\documentclass[runningheads,a4paper]{article}

\usepackage{cmap}
\usepackage{latexsym}
\usepackage[cp1251]{inputenc}
\usepackage[unicode]{hyperref}

\usepackage{amssymb}
\usepackage{graphicx}

\usepackage{tikz}
\usetikzlibrary{shapes,arrows, positioning}
\pgfmathtruncatemacro\distance{1}

\usepackage{url}
\urldef{\mailsa}\path|amironov66@gmail.com|
\urldef{\mailsb}\path|anna.kramer, leonie.kunz, christine.reiss, nicole.sator,|
\urldef{\mailsc}\path|erika.siebert-cole, peter.strasser, lncs}@springer.com|

\begin{document}

\def\xor{
\oplus
}

\def\event#1{#1}

\def\simplus{
\approx
}

\def\soch#1#2{\left(\by
{#1}\\
{#2}
\ey\right)}

\def\ite#1#2#3{\left\{\begin{array}{rlll}
[\![{#1}]\!]&{#2}\\
:&{#3}
\ey\right.}

\def\g#1{\left[\!\!\left[
{\def\arraystretch{1}
\begin{array}{lllll}#1\end{array}}
\right]\!\!\right]}

\def\bcy{\begin{array}{ccc}}
\def\pdownl#1{
  \begin{picture}(8,20)
  \put (4,20){\vector(0,-1){30}}
  \put (1,6){\makebox(1,1)[r]{$\scriptstyle #1$}}
  \end{picture} }
  \def\pleft#1{
  \begin{picture}(20,8)
  \put (25,3){\vector(-1,0){30}}
  \put (12,10){\makebox(1,1){$\scriptstyle #1$}}
  \end{picture} }
  \def\pdownr#1{
  \begin{picture}(8,20)
  \put (4,20){\vector(0,-1){30}}
  \put (7,6){\makebox(1,1)[l]{$\scriptstyle #1$}}
  \end{picture} }

\def\dey#1#2{#1 (#2)}
\def\deyc#1#2{#1 \cdot  #2}
\def\bcy{\begin{array}{ccc}}

\def\ral#1{\;\mathop{\longrightarrow}\limits^{#1}\;}
\def\bc{\begin{center}\begin{tabular}{l}}
\def\ec{\end{tabular}\end{center}}
\def\modn#1{\mathop{=}\limits_{#1}}
\def\inn#1{\mathop{\in}\limits_{#1}}
\def\modnop#1{\mathop{#1}\limits_{n}}
\def\rar{\mathop{\in}\limits_{r}}
\def\und#1{\mathop{=}\limits_{#1}}
\def\skobq#1{\langle\!| #1 |\!\rangle}
\def\redeq{\;\mathop{\approx}\limits^{r}\;}
\def\reduc{\;\mathop{\mapsto}\limits^{r}\;}
\def\pt{\;\mathop{+}\limits_{\tau}\;}
\def\sost{\begin{picture}(0,0)\put(0,3){\circle*{4}}
\end{picture}}
\def\sosto{\begin{picture}(0,0)\put(0,0){\circle*{4}}
\end{picture}}
\def\bi{\begin{itemize}}
\def\pa{\,|\,}
\def\oc{\;\mathop{\approx}\limits^{+}\;}
\def\p#1#2{(\;#1\;,\;#2\;)}
\def\mor#1#2#3{\by #1&\pright{#2}&#3\ey}
\def\ei{\end{itemize}}
\def\bn{\begin{enumerate}}
\def\en{\end{enumerate}}
\def\i{\item}
\def\a{\forall\;}

\def\l#1{[#1]}
\def\ll#1{[\![#1]\!]}
\def\lc#1{\langle#1\rangle}
\def\lcc#1#2{{#1}^{#2}}

\def\fm#1{\left[\!\!\left[\begin{array}{lllll}#1\end{array}\right]\!\!\right]}
\def\ra#1{\mathop{\to}\limits^{\!\!#1}}
\def\hra#1{\mathop{\hookrightarrow}\limits^{#1}}
\def\rb#1{\mathop{\to}\limits_{\!\!#1}}
\def\dra#1{\mathop{\Rightarrow}\limits^{\!\!\!\!#1}}

\def\bigset#1#2{\left\{\by #1 \left| \by #2 \ey\right\}\ey\right.}
\def\p{\leftarrow}

\def\pmiddleright#1{
  \begin{picture}(30,18)
  \put (-5,3){\vector(1,0){40}}
  \put (12,8){\makebox(1,1){$\scriptstyle #1$}}
  \end{picture} }

\def\plongright#1{
  \begin{picture}(40,18)
  \put (-5,3){\vector(1,0){50}}
  \put (20,8){\makebox(1,1){$\scriptstyle #1$}}
  \end{picture} }

\def\plongleft#1{
  \begin{picture}(40,8)
  \put (45,3){\vector(-1,0){50}}
  \put (20,8){\makebox(1,1){$\scriptstyle #1$}}
  \end{picture} }

\def\pse#1#2{
  \begin{picture}(40,8)
  \put (-5,-5){\vector(2,-1){50}}
  \put (45,-5){\vector(-2,-1){50}}
  \put (-5,-12){\makebox(1,1)[r]{$\scriptstyle #1$}}
  \put (45,-12){\makebox(1,1)[l]{$\scriptstyle #2$}}
  \end{picture} }

\def\und#1{\mathop{=}\limits_{#1}}
\def\redeq{\;\mathop{\approx}\limits^{r}\;}
\def\reduc{\;\mathop{\mapsto}\limits^{r}\;}
\def\oc{\mathop{\approx}\limits^{+}}
\def\sost{\begin{picture}(0,0)\put(0,0){\circle*{4}}
\end{picture}}
\def\bi{\begin{itemize}}
\def\pa{\,|\,}
\def\oo{\;\mathop{\approx}\limits^{c}\;}
\def\p#1#2{(\;#1\;,\;#2\;)}
\def\mor#1#2#3{\by #1&\pright{#2}&#3\ey}
\def\ei{\end{itemize}}
\def\bn{\begin{enumerate}}
\def\en{\end{enumerate}}
\def\i{\item}
\def\bigset#1#2{\left\{\by #1 \left| \by #2 \ey\right\}\ey\right.}
\def\p{\leftarrow}
\def\buffer{{\it Buffer}}
\def\eam{\mathbin{{\mathop{=}\limits^{\mbox{\scriptsize def}}}}}
\def\be#1{\begin{equation}\label{#1}}
\def\ee{\end{equation}}
\def\re#1{(\ref{#1})}

\def\bn{\begin{enumerate}}
\def\en{\end{enumerate}}
\def\bi{\begin{itemize}}
\def\ei{\end{itemize}}
\def\i{\item}
\def\c#1{
{\def\arraystretch{1}
\left\{
\begin{array}{lllll}#1\end{array}\right\}}}
\def\d#1{\left[\begin{array}{lllll}#1\end{array}\right]}
\def\b#1{\left(\begin{array}{lllll}#1\end{array}\right)}
\def\ra#1{\;\mathop{\to}\limits^{\!\!#1}\;}
\def\leqd{\;\mathop{<}\limits_{2}\;}
\def\diagrw#1{{
  \def\normalbaselines{\baselineskip20pt \lineskip3pt \lineskiplimit3pt }
  \matrix{#1}}}

\def\blackbox{\vrule height 7pt width 7pt depth 0pt}
\def\pu#1#2{
\mbox{$\!\!\begin{picture}(0,0)
\put (-#1,-#2){\line(1,0){#1}}
\put (-#1,-#2){\line(0,1){#2}}
\put (#1,#2){\line(-1,0){#1}}
\put (#1,#2){\line(0,-1){#2}}
\put (-#1,#2){\line(1,0){#1}}
\put (-#1,#2){\line(0,-1){#2}}
\put (#1,-#2){\line(-1,0){#1}}
\put (#1,-#2){\line(0,1){#2}}
\end{picture}$}
}

\def\pright#1{
  \begin{picture}(20,18)
  \put (-5,3){\vector(1,0){30}}
  \put (9,10){\makebox(1,1){$\scriptstyle #1$}}
  \end{picture} }

\def\by{\begin{array}{llllllllllllll}}
\def\ey{\end{array}}


\title{State diagrams of functional programs
}

\author{Andrew M. Mironov
}

\date{
Moscow State University
\\
Faculty of Mechanics 
and Mathematics\\$\;$\\
amironov66@gmail.com
}


\maketitle

\newcounter{theorem}
\newcounter{lemma}
\newcounter{fig}

\begin{abstract}

In the paper we introduce graphical objects (called state diagrams) 
related to functional programs.
It is shown that state diagrams of 
functional programs can be used  
to solve problems of verification of 
functional programs. 
The proposed approach is 
illustrated by an example 
of verification of a sorting program.

\end{abstract}

\section{Introduction}

The problem of
{\bf program verification}
consists of proving statements 
that analyzed programs 
have specified properties.
This problem
is one of the main problems 
of theoretical computer science.

For various classes of programs there are used various verification methods. 
For example, for a verification of sequential programs there are
used FloydТs inductive assertions method \cite{floyd}
Hoare's logic \cite{hoare},
etc., are used.
For verification
of parallel and distributed programs there are used methods based on 
Milner's  calculus
of communicating systems (CCS) and 
$\pi$--calculus 
\cite{milner1}
 \cite{milner2}, 
Hoare's 
theory of communicating sequential processes (CSP) and its generalizations 
\cite{csp},
\cite{sep},
temporal logic and model checking 
\cite{peled}, 
process algebra
\cite{pa}, Petri nets \cite{petri}, 
etc. are used.
Main methods of verification of functional programs (FPs) are computational
induction and structural induction 
\cite{mironov}. 
Disadvantages of these methods are
related to difficulties to construct formal proofs of program correctness. Among
other methods of verification of FPs it should be noted a method based on
reasoning with datatypes and abstract interpretation through type inference 
\cite{rybal},
a model checking method to verify FPs 
\cite{8},
\cite{14},
methods based on flow analysis
\cite{5},
methods based on the concept of a multiparametric tree transducer 
\cite{9}.

In this article we consider FPs as systems of algebraic equations over strings.
We introduce a concept of a state diagram for such FPs and present the verification method based on  state diagrams. The main advantage of our approach
in comparison with all the above approaches to verification of FPs 
is that
our approach allows to present proofs of correctness of FPs 
in
the form of simple properties 
of their state diagrams.

The basic idea of our approach 
is the following:
\bi
\i we assume that a specification of properties of FP $\Sigma$
under verification
is expressed by another FP
$\Sigma'$, 
whose input is equal to the output 
of  
FP $\Sigma$,
\i we say that a FP $\Sigma$  is correct with respect to the
specification
$\Sigma'$, iff 
the composition 
$f_{\Sigma'}(f_{\Sigma})$  
of input-output maps corresponded to
FPs
$\Sigma$ and  $\Sigma'$
has an output value 1 on all its input values, we denote this statement by the notation
\be{zdfgdfsgdfsgdfs}
f_{\Sigma'}(f_{\Sigma})=1
\ee
\i  we reduce the problem
of a proving  statement
\re{zdfgdfsgdfsgdfs}
to the problem of an analysis of a state
diagram for the FP 
$\Sigma'(\Sigma)$,
whose input-output map
$f_{\Sigma'(\Sigma)}$
is equal 
to the composition
$f_{\Sigma'}(f_{\Sigma})$.\ei

The proposed method of verification of FPs is illustrated by an example of
verification of a sorting FP:
\bi\i at first, we present a  proof of correctness of
this FP by structural induction, 
\i at second, we present a correctness proof of the FP by the method based on
constructing of state diagrams, the proof by the second method 
can be generated automatically. \ei

\section{Main concepts} 

\subsection{Terms} 

We assume that there is given the 
set
${\cal D}$ of {\bf values}, and each 
element of 
${\cal D}$
has one of the following types:
{\bf C},
{\bf S}
or {\bf B}.
The sets of values of the types
{\bf C},  {\bf S}
and {\bf B} are denoted by
 ${\cal D}_{\bf C}$, 
  ${\cal D}_{\bf S}$
and  ${\cal D}_{\bf B}$, respectively,
and
\bi\i
values of the type 
{\bf C} are called {\bf symbols},
\i values of the type 
{\bf S} are called  
{\bf symbolic strings} (or briefly
{\bf strings}),
each string 
 is a finite (maybe empty) sequence of symbols,
\i
values of the type  {\bf B}  are called  
{\bf boolean values},
there are two boolean values:
$\top$ (true) and $\bot$ (false).
\ei


We assume also that there are sets
\bi
\i
${\cal X}$ of {\bf data variables} 
(or briefly {\bf variables}),
\i ${\cal C}$ of {\bf constants},
\i ${\cal F}$ {\bf functional symbols (FSs)}, and
\i $\Phi$ of {\bf functional variables}.
\ei

Each element $x$ 
of any of the above sets
is associated with a {\bf type} 
of this element, 
denoted by the notation
$\tau(x)$, and
\bi
\i if $x\in
{\cal X}$
or $x\in
{\cal C}$,
then $\tau (x)\in \{{\bf C}, {\bf S}, 
{\bf B}\}$, and
\i if $x\in{\cal F}$
or 
$x\in\Phi$, then 
$\tau(x)$ is a notation of the form
$(t_1,\ldots,t_n)\to t$,  where 
$t_1,\ldots, t_n, t \in \{{\bf C}, {\bf S},
{\bf B}\}$.
\ei

Each constant $c\in {\cal C}$
corresponds to an element of
the set
${\cal D}_{\tau(c)}$, called a 
{\bf value} of
this constant.
The notation $\varepsilon$ 
denotes a constant of the type
{\bf S}, whose value is
an empty string.
There are constants 
of the type ${\bf B}$ 
which correspond to the values 
$\top$ and $\bot$, these constants 
are denoted by 
$\top$ and $\bot$ respectively.

Each FS 
$f\in {\cal F}$ 
corresponds to a partial function, which is denoted 
by the same symbol 
$f$, and has the form
$$f: {\cal D}_{t_1}\times\ldots \times
{\cal D}_{t_n} \to {\cal D}_t,\quad
\mbox{ where }\;
\tau(f)=
(t_1,\ldots,t_n)\to t.$$

Below we list some of the FSs 
which belong to ${\cal F}$,
beside each FS we point
out (with a colon) its type.
\bn
\i $head: {\bf S} \to {\bf C}$.
   The function $head$ 
   is defined for non-empty strings, it maps
each non-empty string to its first element (i.e. if a string $u$ 
has the form
$a_1\ldots a_n$, then
 $head(u)=a_1$).
\i $tail: {\bf S} \to {\bf S}$.
   The function $tail$ 
   is defined for non-empty strings, it maps each
non-empty string $u$ 
u to a string
(called a 
{\bf tail} of the string $u$),
  derived from
  $u$ by removal
of its first element
 (i.e. if a string $u$ 
has the form
$a_1a_2\ldots a_n$, then
 $tail(u)=a_2\ldots a_n$).
\i $conc: ({\bf C}, {\bf S})\to {\bf S}$.
   For each pair
   $(a,u)\in {\cal D}_{\bf C}\times 
   {\cal D}_{\bf S}$ 
   S the string
    $conc(a,u)$ is derived 
by a writing the symbol
   $a$ before $u$.

\i $= :(t,t)\to {\bf B}$,
 where $t\in\{{\bf C},{\bf S},{\bf B}\}$,
i.e. the symbol $=$ 
denotes three FSs.
A value of the function   
$=$ on the pair
$(x, y)$ is $\top$, if 
$x$ and $y$ are equal,
and $\bot$, otherwise.

\i $\leq :({\bf C}, {\bf C})\to {\bf B}$.
We assume that
${\cal D}_{\bf C}$ is a
linearly ordered set, and 
 the value
of the function 
$\leq$ on the pair
$(a, b)$ is
$\top$, if $a\leq b$, and $\bot$, otherwise.

\i Boolean FSs:
$$\neg: {\bf B}\to {\bf B},\quad
\wedge: ({\bf B},{\bf B})\to {\bf B},\quad
\mbox{etc.},$$ the corresponding functions are
standard boolean functions on the arguments
$\top$ and $\bot$
(i.e. $\neg (\top) = \bot$, etc.).

\i ${\it if\_then\_else}\;: ({\bf B},t,
    t)\to t$, where $t\in\{{\bf C},{\bf S},{\bf B}\}$,
i.e.  the notation ${\it if\_then\_else}$ 
denotes three FSs.
Functions corresponding to
these FSs  
 are defined as follows:
   $${\it if\_then\_else}\;(a,x,y)\eam
   \left\{\by x,\;\;\mbox{if $a=\top$,}\\
   y,\;\;\mbox{if $a=\bot$}. \ey\right.$$
\en

A concept of a 
{\bf term} 
is defined inductively. Each term
$e$ is associated with a
 type
 $\tau(e) \in \{{\bf C}, {\bf S},
 {\bf B}\}$.
 A definition of a term has the following
 form:
  \bi\i
  each data variable 
  and each constant is a term,
  its type is equal to the type of this variable or constant,
\i if
$f$ is a FS or 
a functional variable,
$e_1,\ldots, e_n$ are terms, and
$$\tau(f) = (\tau(e_1),\ldots, \tau(e_n))\to t,$$ 
then 
$f(e_1,\ldots, e_n)$ 
is a term of the type
$t$.\ei

We shall use the following
concepts and notations.
\bi
  \i A set of all terms is denoted by the
  symbol
      ${\cal E}$.
      \i Terms of the type ${\bf B}$ are called
       {\bf formulas}.
\i     $\forall\,e,e'\in {\cal E}$
	$e'$ is a {\bf subterm}
	of $e$, if either
	$e'=e$, or
	$e=f(e_1,\ldots,e_n)$, and $\exists\,i\in
	\{1,\ldots, n\}:e'$ is a subterm 
of  $e_i$.
\i     $\forall\,e\in {\cal E}\;\;
X_{e}$ and $\Phi_{e}$ are
      sets of data variables and
      functional variables 
      respectively,
occurred in $e$.
\i $\forall\,X\subseteq {\cal X}\;\;
{\cal E}_X\eam
\{e\in {\cal E}\mid
X_e\subseteq X\}$.
\i
The terms
$$\by
head(e),\; tail(e),\;conc(e, e'),
    =(e,e'),\;
\leq (e,e'),\;
       {\it if\_then\_else}\;(e,e',e'')\ey$$
       are denoted by 
   $e_h$, $e_t$, $e e'$, 
         $e = e'$,
         $e \leq e'$,
      $[\![e]\!]\,e':e''$,
respectively.
\i A term $e\in {\cal E}$ is said to be 
{\bf simple}, if $e=e_1\ldots e_n$, 
 where each term from the list
$e_1$, $\ldots$, $e_n$ is 
a data variable
or a constant.
\i    Terms containing boolean FSs will be denoted as in mathematical
texts (i.e. in the form 
      $e\wedge e'$, etc.),
terms of the form
$e_1\wedge\ldots\wedge e_n$
can also be denoted 
by the notation
 $
{\def\arraystretch{0.5}
\left\{\by 
e_1\\\ldots\\e_n\ey\right\}}$,
\i
$\forall\,e\in {\cal E}$
the notation ${\cal D}_e$ 
denotes the set 
${\cal D}_{\tau(e)}$.
\i Lists of terms are denoted by 
the notations of the form
$\bar e$.
\i
If $\bar e$ is a list of terms
of  the form
$(e_1,\ldots, e_n)$, 
then \bi\i 
$\tau(\bar e)$ denotes the list
$(\tau(e_1)$,
$\ldots$, $\tau(e_n))$,
\i  
$X_{\bar e}$, $\Phi_{\bar e}$
denote the sets
$\bigcup_{i=1}^n X_{e_i}$, 
$\bigcup_{i=1}^n \Phi_{e_i}$
respectively,
\i  ${\cal D}_{\bar e}$
denotes the set
${\cal D}_{e_1}\times
\ldots\times {\cal D}_{e_n}$.
\ei

\i
If $\bar e'=(e'_1,\ldots, e'_n)$,
$\bar e''=(e''_1,\ldots, e''_n)$ are
lists of terms, 
$\tau(\bar e')=
\tau(\bar e'')$, 
then \bi\i
the notation
$\bar e'=\bar e''$
denotes the term
$(e'_1=e''_1)\wedge\ldots
\wedge (e'_n=e''_n)$,
\i if in addition
it is assumed that for each pair
$i,j$ of different indices from 
$\{1,\ldots, n\}$
the term $e'_i$ is a subterm of 
$e''_j$,
then \bi\i
$\forall\,e\in {\cal E}$
the notation \be{sadfdsgfdsgsew3}
e[e''_1/e'_1,\dots, e''_n/e'_n]\ee
denotes a term derived from
$e$ by replacing
 $\forall\,i=1,\ldots, n$
 each subterm of 
$e$, 
 which is equal to  $e'_i$,
on the term $e''_i$,
term
\re{sadfdsgfdsgsew3}
is denotes also by the notation
$e[\bar e''/\bar e']$,
\i for each list of terms
$\bar e=(e_1,\ldots, e_m)$ the notation
 $\bar e[\bar e''/\bar e']$
denotes the term
 $$(e_1[\bar e''/\bar e'],\ldots, e_m[\bar e''/\bar e']).$$
\ei
 \ei
 \i A {\bf clarification}
 is a notation 
$\theta$ of the form
\be{sdffgdfsgsdfgsdfgfds}
 e_1/x_1,\ldots, e_n/x_n,\ee
 where $x_1,\ldots, x_n$ are 
 different variables,
 $e_1,\ldots, e_n$ are simple terms,
such that 
$\forall\,i=1,\ldots, n\;\;\tau(x_i)=\tau(e_i)$.
  $\forall\,e\in {\cal E}$
the notation
$e[\theta]$ denotes the term
 $e[e_1/x_1,\ldots, e_n/x_n]$
 (similar notations are used when a list of terms is considered instead of the a term $e$).
 
 \re{sdffgdfsgsdfgsdfgfds}
is called a {\bf renaming}, if $e_1,\ldots, e_n$ are different variables.
\ei

\subsection{A concept of a functional program}

In this article, a 
{\bf functional program (FP)} 
 refers to a finite set 
 $\Sigma$
 of equalities of the form
\be{dfgdsfgdsr5rtt6}
\left\{\by\varphi_1(x_{11},\ldots, x_{1n_1})=e_1\\
\ldots\\
\varphi_m(x_{m1},\ldots, x_{mn_m})=e_m
\ey\right.\ee
where\bi\i
$\varphi_1, \ldots, \varphi_m$ are
different 
functional variables, and
\i $\forall\,i=1,\ldots, m$
$\varphi_i(x_{i1},\ldots,x_{in_i})$
      and $e_i$ are terms of the same 
      type, and
  $$X_{e_i}=\{x_{i1},\ldots,x_{in_i}\},\quad
  \Phi_{e_i}\subseteq\{\varphi_1,\ldots,
\varphi_m\}.$$\ei

A {\bf main term} of 
FP \re{dfgdsfgdsr5rtt6}
is the left side of  first equality in
\re{dfgdsfgdsr5rtt6} (i.e. the term
$\varphi_1(x_{11},\ldots, x_{1n_1})$).

The set of equalities in FP  \re{dfgdsfgdsr5rtt6}
can be considered as a system of functional equations for functional variables
$\varphi_1,\ldots, \varphi_m$.
This system defines a list
\be{dfasdfasd}(f_{\varphi_1},\ldots, f_{\varphi_m})\ee
of partial functions corresponding to $\varphi_1$,
$\ldots$, $\varphi_m$,
which is the least (in the sense of the order on lists of partial functions described in
\cite{mironov})
a solution of system of functional equations
\re{dfgdsfgdsr5rtt6}.
List \re{dfasdfasd}
is called a
 {\bf least fixpoint (LFP)} of FP 
\re{dfgdsfgdsr5rtt6}.
All details related to the concept of a LFP of a FP,
can be found in chapter 5 of the
 book \cite{mironov}.
The first function in the list
\re{dfasdfasd}
(i.e.  $f_{\varphi_1}$) is denoted by
$f_\Sigma$,  and is called
a {\bf function defined by the FP
$\Sigma$}.

Let $\Sigma$ be a FP.
The notation ${\cal E}_\Sigma$
denotes the set of all terms, such that all  functional variables occurred
in them, are occurred in $\Sigma$. 

FPs $\Sigma$ and $\Sigma'$
are considered as equal, if 
$\Sigma'$ is derived from  $\Sigma$
by renaming of data variables 
and functional variables, i.e. if $X\Phi_\Sigma$
and $X\Phi_{\Sigma'}$ are sets of 
data variables 
and functional variables occurred in
$\Sigma$
and $\Sigma'$ respectively, then
 there is a one-to-one correspondence $f:X\Phi_\Sigma\to 
X\Phi_{\Sigma'}$, such that
$\Sigma'$ is derived from $\Sigma$ 
by replacing each variable 
$v\in X\Phi_\Sigma$ on $f(v)$.

\section{An example of 
specification and verification  of a 
functional program}

\subsection{An example of 
a  
functional program
}
\label{4.3.2}
\label{zsdfgdfsgdsfgdfsgsdffgds1}


Consider the following FP:
\be{fdsgdsgdsr}
\left\{\by
\by {\bf sort}(x) = [\![x = \varepsilon]\!] \,\varepsilon
 :  {\bf insert}(x_h, {\bf sort}(x_t))\ey\\
\begin{array}{rlll}  {\bf insert}(a, y) = [\![y = \varepsilon]\!]& a  \varepsilon\\
:&[\![a \leq  y_h]\!] & a  y\\
&&: y_h   \;{\bf insert}(a, y_t)\ey\ey\right.\ee
This FP defines a sorting function on strings. The FP consists of two equations that define the following functions:
\bi
\i ${\bf sort}: {\bf S} \to {\bf S}$ is
a main function,  and
\i ${\bf insert}: ({\bf C}, {\bf S})\to {\bf S}$ is an auxiliary function,
this function  maps a pair
$(a,y)\in {\cal D}_{\bf C}\times
 {\cal D}_{\bf S}$ 
to the string derived by inserting a character $a$ to the string 
$y$, such that the following condition
holds: if the string
$y$ is ordered, then 
the string ${\bf insert}(a,y)$ also is 
ordered
(a string is ordered, if its 
components form a non-decreasing sequence). 
\ei

\subsection{An example of a specification of a functional program
} \label{sdfsadfsad}

One of the properties of correctness
of  FP \re{fdsgdsgdsr}
has the form:
$\forall\,x\in {\cal D}_{\bf S}$ 
the string ${\bf sort}(x)$ is ordered.
This property can be described formally as follows.
Consider
a FP defining a function
{\bf ord}  of string ordering checking:
\be{sdfgfdsgsdfgdsfgsrrr}
\by
{\bf ord}(x) 
=[\![x = \varepsilon]\!] &1\\
&: [\![x_t = \varepsilon]\!] &1\\
&&: [\![x_h \leq (x_t)_h]\!] &{\bf ord}(x_t): 0\ey\ee

The function {\bf ord} allows to describe the above property of correctness as the
following statement:
\be{4.21}\forall\,  x \in 
 {\cal D}_{\bf S}\quad
{\bf ord}({\bf sort}(x)) = 1. \ee

\subsection{An example of verification
of a functional program}
\label{zsdfgdfsgdsfgdfsgsdffgds}

The problem of verification 
of the correctness property of FP
\re{4.21} of FP \re{fdsgdsgdsr}
consists of a
formal proof of proposition
\re{4.21}.
This proposition can be proved like an ordinary mathematical
theorem, for example using the method of mathematical induction. 
A proof of this proposition can be the following.

If $x = \varepsilon$, then, according to first equation of system
\re{fdsgdsgdsr},  the equality
 ${\bf sort}(x) = \varepsilon$ holds, and therefore
$${\bf ord}({\bf sort}(x)) = {\bf ord}(\varepsilon) = 1.$$

Let $x \neq  \varepsilon$.
We prove  \re{4.21} 
for this case by induction. Assume that for each
string $y$, which is shorter than $x$, the equality
$${\bf ord}({\bf sort}(y)) = 1$$
holds.  Prove that this implies the equality
\be{dfasdfs}{\bf ord}({\bf sort}(x)) = 1.\ee

\re{dfasdfs} is equivalent to the equality
\be{4.22}
{\bf ord}( \;{\bf insert}(x_h, {\bf sort}(x_t))) = 1. \ee
By the induction hypothesis, the equality $${\bf ord}({\bf sort}(x_t)) = 1,$$
holds, and this implies
\re{4.22} on the reason of the following lemma. \\

{\bf Lemma}.

The following implication holds:\be{4.23}
{\bf ord}(y) = 1 \quad\Rightarrow\quad
 {\bf ord}( {\bf insert}(a,  y)) = 1 \ee

{\bf Proof}.

We prove the lemma by induction on the length of
$y$.

If $y = \varepsilon$,  then the right side of 
\re{4.23} has the form
$${\bf ord}(a  \varepsilon) = 1,$$
which is true by definition of
{\bf ord}.

Let $y \neq  \varepsilon$,  and for each string $z$, which is shorter than
$y$, the following
implication holds:
\be{4.24}
{\bf ord}(z) = 1 \quad\Rightarrow\quad 
{\bf ord}( {\bf insert}(a, z)) = 1 \ee

Let $c \eam  y_h$, $d \eam y_t$.
Then
\re{4.23} has the form
\be{4.25}
{\bf ord}(c  d) = 1 \quad
\Rightarrow\quad
 {\bf ord}( {\bf insert}(a,  c   d)) = 1\ee

To prove the implication
\re{4.25}
 it is necessary to prove that if 
 ${\bf ord}(c  d) = 1$, then
  the following implications hold:
\bi
\i[(a)] $a \leq c \quad\Rightarrow\quad 
{\bf ord}(a   (c   d)) = 1$,
\i[(b)] $c < a \quad\Rightarrow\quad  {\bf ord}(c   
 \;{\bf insert}(a,  d)) = 1$.
\ei

(a) holds because 
$a \leq c$ implies
$${\bf ord}(a   (c   d)) = {\bf ord}(c   
d) = 1.$$

Let us prove (b).

\bi
\i $d = \varepsilon$. 
In this case, right side of (b) 
has the form
\be{4.26}
{\bf ord}(c   (a   \varepsilon)) = 1\ee

\re{4.26} follows from
$c <a$.
\i $d \neq  \varepsilon$. 
Let $p \eam  d_h$, 
$q \eam  d_t$.

In this case, it is necessary to prove that if
$c < a$, then
\be{4.27}
{\bf ord}(c     \;{\bf insert}(a, p   q)) = 1 \ee

\bn
\i if $a \leq p$, then
\re{4.27} has the form
\be{4.28}
{\bf ord}(c    (a   (p   q))) = 1\ee

Since $c < a\leq  p$, then \re{4.28}
follows from the equalities
$$\by
{\bf ord}(c   (a   (p   q))) = {\bf ord}(a   (p   q)) = {\bf ord}(p   q) =\\
= {\bf ord}(c   (p   q)) = {\bf ord}(c   d) = 1\ey$$
\i if $p <a$, then \re{4.27} has the form
\be{4.29}
{\bf ord}(c   (p    \;{\bf insert}(a, q))) = 1 \ee

Since, by assumption,
$${\bf ord}(c   d) = {\bf ord}(c   (p   q)) = 1$$
then $c \leq p$, 
 and therefore
 \re{4.29} 
can be rewritten as
\be{4.30}
{\bf ord}(p    \;{\bf insert}(a, q)) = 1\ee 

If $p <a$, then
$$ {\bf insert}(a, d) =  {\bf insert}(a, p   q) 
= p    \;{\bf insert}(a, q)$$
therefore \re{4.30} 
can be rewritten as
\be{4.31}{\bf ord}( {\bf insert}(a, d)) = 1\ee 
\re{4.31}  follows from the induction hypothesis for the Lemma (i.e., from the
implication 
\re{4.24},
 where $z \eam d$) 
and from the equality
$${\bf ord} (d) = 1$$
which is justified by the chain of equalities
$$\by 1 = {\bf ord}(c   d) = {\bf ord}(c   
(p    q)) = \quad(\mbox{since }c \leq p)\\
= {\bf ord}(p   q) = {\bf ord}(d).\quad\blackbox\ey
$$
\en
\ei

From the above example we  see that even for the simplest FP, which
consists of several lines, 
\bi\i
a proof of its correctness is not trivial mathematical
reasoning, \i it is difficult to check this proof, and \i it is
much more difficult to construct this proof.\ei
Below we present a radically different method for verification of FPs based
on a construction of state diagrams for FPs. We illustrate our approach
by a proof of 
the proposition \re{4.21}
on the base of the proposed method. This proof can be generated automatically, that is
an evidence of an advantage of the method for verification of FPs based on state diagrams.

\section{States of functional 
programs}
\label{sadfgdsgdfsgdsfgds}

\subsection{A concept of a state
of a functional program}

Let $\Sigma$ be a FP.
A {\bf state} of $\Sigma$ 
is a notation $s$ of the form
$\{\beta\}^y_{x_1,\ldots, x_n}$,
where
\bi
\i $\beta\in {\cal E}_\Sigma$ is a formula, 
called a {\bf condition} of the state $s$,
\i $y$ is a simple term, called an {\bf output term} of the state $s$,
and
\i $x_1,\ldots, x_n$ is a list of 
simple terms, called
{\bf input terms} of the state 
$s$.\ei

We shall use the following notations:
\bi\i
the set of all states of FP  
$\Sigma$  
 is denoted by 
${\cal S}_\Sigma$, \i
$\forall\,s\in {\cal S}_\Sigma$ 
 the set of all data variables 
 occurred in $s$ is denoted by
$X_s$,\i
if the state $s$ 
has the form 
$\{\beta\}^y_{x_1,\ldots, x_n}$, then
the terms $\beta$, $y$ and
the list
$x_1,\ldots, x_n$ can be denoted by
$\beta_s$, $y_s$, and $\bar x_s$, 
respectively,\i
if $\beta_s$ has the form
$e_1\wedge \ldots\wedge e_n$,
then $s$ can  be denoted by
 $
{\def\arraystretch{0.5}
\left\{\by 
e_1\\\ldots\\
 e_n\ey\right\}^{y_s}_{\bar x_s}}$,\i
 if  $\beta_s=\top$, 
 then 
$s$ can  be denoted by
$\{\}^{y_s}_{\bar x_s}$.\ei

A state $s\in {\cal S}_\Sigma$
is said to be an {\bf initial} 
state of FP $\Sigma$
(and is denoted by
$s^0_\Sigma$), 
if it has the form 
$\{y=\varphi(\bar x)\}^y_{\bar x}$
 where \bi\i
$\varphi$ is a functional 
variable occurred in main term 
of FP $\Sigma$,
\i $\bar x$ is a list of different 
variables, \i
 $y$ is a variable which is not occurred in 
$\bar x$, and \i
$\tau(\varphi)=
\tau(\bar x)\to \tau(y)$.\ei

A state $s\in {\cal S}_\Sigma$
is said to be 
{\bf terminal}, 
if $\Phi_{\beta_s}=\emptyset$. 

If $s\in {\cal S}_\Sigma$
and $\theta$ is a clarification, 
then $$s[\theta]\eam
\{\beta_s[\theta]\}^{y_s[\theta]}_{\bar x_s[\theta]}.$$
If  $\theta$ is a renaming, then 
we say that 
$s[\theta]$ is derived from 
$s$ by a renaming.

\subsection{Equality of terms and 
states}

Let $X$ be a subset of ${\cal X}$.
An {\bf evaluation} of variables
from $X$
is a function $\xi: X\to{\cal D}$,
which maps each variable 
$x\in X$ to a value 
$\xi(x)$ of the type $\tau(x)$. 
A set of all  evaluations of variables
from $X$
$X$ is denoted by $X^\bullet$.
For 
\bi\i each evaluation 
$\xi\in X^\bullet$,
and \i each term $e$,  
such that
 $X_e\subseteq X$
and $\Phi_e=\emptyset$,\ei
 $e^\xi$ denoted an object which either is a value from
 ${\cal D}$, or is not defined,
and is defined, and is computed recursively:
\bi
\i if $e=x\in X$, then $e^\xi=\xi(x)$,
\i if $e=c\in {\cal C}$, then $e^\xi$ 
is a value of the constant
$c$,
\i if $e=f(e_1,\ldots, e_n)$, where $f\in {\cal F}$,
then
\bi\i $e^\xi$  is equal to the value
$f(e_1^\xi, \ldots, e_n^\xi)$, if this value is defined, \i  
$e^\xi$ is not defined, otherwise.\ei
\ei

Let $\Sigma$ be a FP. 
$\forall\,e\in {\cal E}_\Sigma$, 
$\forall\,\xi\in X^\bullet:
X\supseteq X_e$, the notation
$e^{\xi,\Sigma}$ denotes a
{\bf value of $e$ on $\xi$ with respect to $\Sigma$},
which is defined as above, 
but with the following difference:
functional variables
from 
$\Phi_e$ are considered as FSs, 
associated with partial functions, 
which are corresponding components of a least fixpoint of $\Sigma$.

Terms $e_1$ and $e_2\in {\cal E}_\Sigma$ 
are considered as {\bf equal} (with respect to the FP $\Sigma$),
if 
$\forall\,\xi\in (X_{e_1}\cup X_{e_2})^\bullet$
the objects $e_1^{\xi,\Sigma}$ and $e_2^{\xi,\Sigma}$ are both
\bi
\i either not 
defined,
\i or defined and 
equal.\ei

Examples of pairs of 
equal terms:
$$\by\;
e_1e_1'=e_2e_2'\;\mbox{ and }\;
(e_1=e_2)\wedge (e'_1=e'_2),\\
\;ee'=\varepsilon\;\mbox{ and }\; \bot,
\\
\;[\![\top]\!]e:e'\;\mbox{ and }\;  e,\\
\;[\![\bot]\!]e:e'\;\mbox{ and }\;  e',\\
\;e=\top\;\mbox{ and }\; e,\\
\;e=\bot\;\mbox{ and }\; \neg e,\\
\;e\wedge (e'=e'')\;\mbox{ and }\;
e[e''/e']\wedge(e'=e'').
\ey$$

Let $\Sigma$ be a FP.
States 
$s,s'\in
{\cal S}_\Sigma$ are considered as 
equal, if one of the following 
conditions hold:
\bi\i
$s'$ can be derived from 
$s$ by a renaming, and 
$\bar x_s=\bar x_{s'}$,
\i 
$\beta_{s'}=
\beta \wedge
(x=e)$,
 where  $x\in {\cal X}$, 
$x\not\in X_s$, 
$e\in {\cal E}$, 
$\beta$ can be derived from 
$\beta_s$ by a replacement of 
some occurrences of 
$e$ on $x$,
$y_{s'}=y_{s}$,
$\bar x_{s'}=\bar x_{s},$
\i 
$\beta_{s}=\beta\wedge (x=e)$, 
where
$x\in {\cal X}$,
 $e$ is a simple term, 
 $x\not\in X_e$,
 $s'=s[e/x]$.
\ei 

\subsection{Transitions 
of functional programs}
\label{perehody}

In this section we define a concept
 of a transition of a  FP. 
A transition of a FP 
\bi\i represents a relation between 
states of the FP, and 
has a {\bf label} which is
\bi
\i either a functional variable, or 
\i or a formula,
called a {\bf condition}
of this transition.
\ei
\ei

A {\bf transition} of 
FP $\Sigma$ is a triple
$r=(s,s',l)$, where 
$s,s'$ are states from 
${\cal S}_\Sigma$, called
a {\bf start} and an {\bf end} of the transition $r$, respectively,
and $l$ is a 
{\bf label} of 
transition $r$.
A transition
$(s,s',l)$ is called a 
transition from $s$ to $s'$.
It can be denoted by 
$s\ra{l} s'$.

Let  $s=\{\beta\}^y_{\bar x}
\in
{\cal S}_\Sigma$.
There are the following 
transitions starting from $s$:
\bn
\i
if 
$\beta$ contains a subterm
of the form
$\varphi(\bar e)$,
and one of equations in
$\Sigma$
has the form
$\varphi(\bar x)=e,$ then there is a 
transition
(called an {\bf expansion})
$$\by s&\pright{\varphi}&\{
\beta[e[\bar e/\bar x]/\varphi(\bar e)]
\}^y_{\bar x},\ey$$
\i if $\beta$ contains a subformula $e$, 
then there is a pair of transitions
\be{compofgdsgsd1}
\by
s&\pright{e}&\{\beta \wedge e\}^y_{\bar x},\quad
s&\pright{\neg e}&\{\beta \wedge \neg e\}^y_{\bar x},
\ey\ee
\i if $\beta$ contains a subterm  $e$ of the type ${\bf S}$, 
then there is a pair 
of transitions
\be{compofgdsgsd2}
\by
s&\pright{e=\varepsilon}&\{\beta
\wedge (e=\varepsilon)\}^y_{\bar x},\quad
s&\pright{e=xx'}&\{\beta
 \wedge(e=xx')\}^y_{\bar x},
\ey\ee
where
$x,x'$ are
{\bf fresh} variables
(which are not occurred in $X_{s}$).
\en

Any transition, occurred  
in a pair of the form 
\re{compofgdsgsd1} or
\re{compofgdsgsd2}, 
is said to be {\bf complementary}
to another transition from this pair.

A set of all 
 transitions of 
  FP $\Sigma$
  is denoted by 
${\cal R}_\Sigma$.

\section{Neighborhoods 
of states of functional programs}

\subsection{Unfoldings of states}

Let $\Sigma$ be a FP.

An
{\bf unfolding} of a state
$s\in {\cal S}_\Sigma$
is a finite tree
$V$, \bi
\i each node $v$ of which is 
associated with a state
$s_v\in 
{\cal S}_\Sigma$,  
\i a root of this tree 
is associated with the state
$s$,  and
\i each edge $r$ of which
is associated with a 
transition 
from ${\cal R}_\Sigma$
of the form
$s_v\ra{l}s_{v'}$,
 where $v$ and $v'$ 
are a start and an end of the edge
$r$, and a label
$l$ of this transition is also 
a label of the edge $r$.
 \ei

Nodes and edges of  $V$ 
will be identified with those states and transitions respectively, 
which are associated with them.

\subsection{A concept of a neighborhood of a state}

Let $\Sigma$ be a FP. 
Each state
$s\in {\cal S}_\Sigma$
is associated with a set
${\cal U}_s$
of 
{\bf 
neighborhoods} of the state $s$.
Each neighborhood $U\in 
{\cal U}_s$  is a tree,
\bi\i nodes of which are associated with states from 
${\cal S}_\Sigma$, and \i edges of which are 
labeled by  lists of labels 
used in unfoldings of states.\ei
The set ${\cal U}_s$ 
is defined as follows.
\bn
\i Each unfolding $V$
of $s$,
such that
$\forall\,v\in V$
the set of edges outgoing from $v$
\bi\i either is empty (in this case $v$
is said to be a {\bf leaf}),
\i or consists of only edge, 
which is labeled by an expansion, 
\i or consists of two 
complementary
edges,
\ei
belongs to the set
${\cal U}_s$.
\i Let $U\in {\cal U}_s$, 
and $s'$ is a node of $U$,
which is not 
a root or leaf, then if
\bi
\i an edge ended in $s'$, 
has the form
$s_0\ra{l}s'$, 
and \i edges started in $s'$, 
have the form
$s'\ra{l_1}s_1$, $\ldots$,
$s'\ra{l_n}s_n$,\ei
then ${\cal U}_s$ has a tree
$U'$, derived from by 
$U$\bi\i
a removing of the node 
$s'$ and  edges related 
to this node, and
\i adding edges
$s_0\ra{ll_1}s_1, \ldots,
s_0\ra{ll_n}s_n$,
where
$\forall\,i=1,\ldots, n\;\; 
ll_i$ is a concatenation of lists 
  $l$ and $l_i$.\ei
\i Let $U\in {\cal U}_s$, and 
$s'$ is a {\bf contradictory} 
node $U$ (i.e. 
$\beta_{s'}=\bot$), 
which is not a root,
then 
${\cal U}_s$ has a tree
$U'$,
derived from $U$
by a removing of nodes reachable 
from $s'$ (i.e. such that there are 
paths from 
$s'$ to these nodes),
and edges related to these nodes.
\en

A neighborhood $U'$, derived from 
$U$ according to items
2 and 3 of this definition,
is said to be a 
{\bf reduction} of the
neighborhood $U$.

It is not so difficult to prove that
\bi\i
a node  $s$ of some 
neighborhood is 
contradictory
iff ends of all edges outgoing from $s$
are contradictory, and \i
a state is contradictory iff
all leaves of some its 
neighborhood are contradictory.
\ei

If $U$ is a neighborhood 
of some state, then
$\forall\,
v\in U$ there is a unique path
from a root of $U$ to the node
$v$. All nodes of 
$U$, lying on this path and not coinciding with $v$,
are said to be 
{\bf ancestors} of $v$.

We shall use the following 
agreement in graphical representation 
of neighborhoods:
if $U$ is a neighborhood of some 
state, then in a graphical 
representation of the 
neighborhood
 $U$
\bi\i
nodes of $U$
are represented by ovals, 
\i a root of $U$ is represented by a double oval, 
\i contradictory nodes can be represented by black boxes 
($\blackbox$),
\i
$\forall\,s\in U$
an oval $O_{s}$, representing 
 $s$, has the following form:
\bi
\i 
conjunctive terms occurred  in 
$\beta_{s}$, are displayed 
in a column inside  
$O_{s}$ (if $\beta_{s}=\top$, then nothing is drawn inside
$O_{s}$),
\i the list  $\bar x_{s}$ of input terms and output term 
$y_{s}$
of the state $s$
are displayed to the right of
$O_{s}$ from the bottom and from the top, respectively, and
\i an identifier of
the state 
$s$ is displayed at the top from the 
left of $O_{s}$,
\ei
\i edges occurred in $U$
are represented by arrows connecting ovals:   
if $U$  contains the edge $s\ra{l}s'$,
then \bi\i
 then this edge is represented by an arrow from $O_{s}$ to
$O_{s'}$, and \i near this arrow the components of the 
 label $l$ may be depicted.
\ei

\ei

\subsection{Examples of 
neighborhoods}

In this section, we give examples of neighborhoods of states for
a FP of sorting \re{fdsgdsgdsr} and for FP
of checking the ordering of strings
\re{sdfgfdsgsdfgdsfgsrrr}.

\subsubsection{Examples of 
neighborhoods for  
the program of sorting}

We rewrite the FP
of sorting
\re{fdsgdsgdsr},  using shorter notation for the function variables occurred in it (we denote terms of the form
${\bf sort}(x)$
and ${\bf insert}(a,y)$
by the notations 
$\varphi(x)$
and $a\to y$  respectively):
\be{fdsgdsgds33r}
\left\{\by
\varphi(x) = [\![x = \varepsilon]\!] \,\varepsilon
 :  (x_h\to \varphi(x_t))\\
a\to y = [\![y = \varepsilon]\!] a  \varepsilon:
\Big([\![a \leq  y_h]\!] \, a  y
: y_h   (a\to y_t)\Big)\ey\right.\ee

One of unfoldings of the state 
$s_0\eam \{y=\varphi(x)\}^y_x$
 consists of the following states and edges:
 $$\by
\by s_0&
\pright{
\varphi
}
&s \eam
\{y=
[\![x = \varepsilon]\!] \,\varepsilon
 :  x_h\to \varphi(x_t)\}^y_x,
 \ey
\\ \by s&
\pright{
x=\varepsilon
}
&\c{y=
[\![x = \varepsilon]\!] \,\varepsilon
 :   x_h\to \varphi(x_t)\\
 x=\varepsilon}^y_x=
 \{y=\varepsilon\}^y_\varepsilon=
\{\}^\varepsilon_\varepsilon,
 \ey
\\ \by s&
\pright{
x=ab
}\quad
\c{y=
[\![x = \varepsilon]\!] \,\varepsilon
 :   x_h\to \varphi(x_t)\\
x=ab}^y_x=\\&\;\hspace{10mm}
=  \{y=a\to \varphi(b)\}^y_{ab}=
  \c{y=a\to p\\
  p=\varphi(b)}^y_{ab}.
\ey
\ey
$$

One of neighborhoods,
corresponded to this unfolding, 
has the form
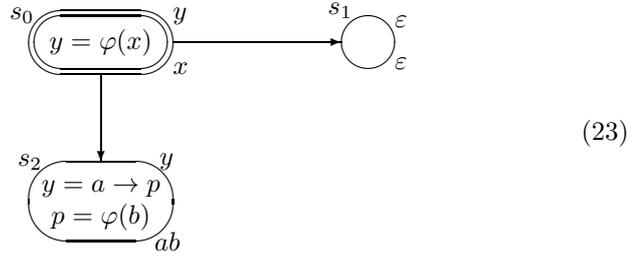
\be{sdafasdfasdf1}
\by
\begin{picture}(50,95)

\put(27,90){\makebox(0,0)[l]{$y$}}
\put(27,70){\makebox(0,0)[l]{$x$}}
\put(0,80){\oval(50,20)}
\put(0,80){\oval(54,24)}
\put(0,80){\makebox(0,0){$
y=\varphi
(x)$}}

\put(110,88){\makebox(0,0)[l]{$\varepsilon$}}
\put(110,72){\makebox(0,0)[l]{$\varepsilon$}}
\put(100,80){\oval(20,20)}

\put(22,35){\makebox(0,0)[l]{$y$}}
\put(20,5){\makebox(0,0)[l]{$ab$}}
\put(0,20){\oval(54,30)}
\put(0,20){\makebox(0,0){$\bcy
y=a\to p\\
p = \varphi(b)
\ey$}}

\put(-25,90){\makebox(0,0)[r]{$s_0$}}
\put(85,92){\makebox(0,0)[l]{$s_1$}}
\put(-22,35){\makebox(0,0)[r]{$s_2$}}

\put(0,68){\vector(0,-1){33}}
\put(27,80){\vector(1,0){63}}

\end{picture}
\ey
\ee

One of 
neighborhoods of 
state $s_2$
in \re{sdafasdfasdf1}
has the form
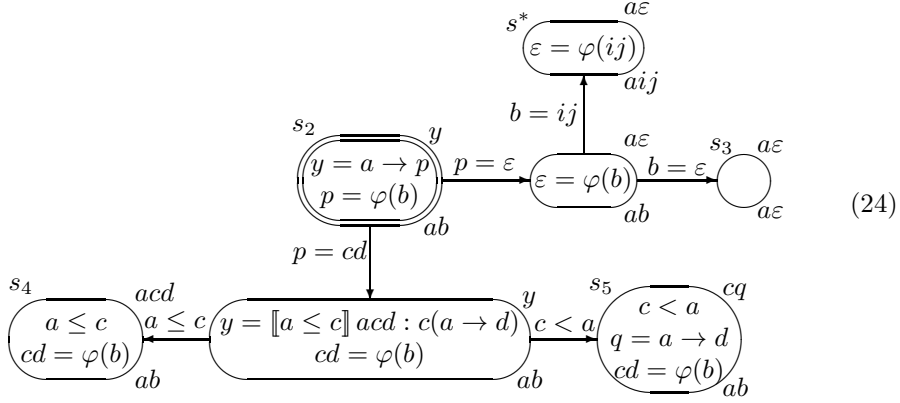
\be{sdafasdfasdf2}
\by
\begin{picture}(50,145)

\put(22,97){\makebox(0,0)[l]{$y$}}
\put(20,63){\makebox(0,0)[l]{$ab$}}
\put(0,80){\oval(50,30)}
\put(0,80){\oval(54,34)}
\put(0,80){\makebox(0,0){$\bcy
y=a\to p\\
p = \varphi(b)
\ey$}}

\put(95,95){\makebox(0,0)[l]{$a\varepsilon$}}
\put(95,68){\makebox(0,0)[l]{$ab$}}
\put(80,80){\makebox(0,0){$\varepsilon=\varphi(b)$}}
\put(80,80){\oval(40,20)}

\put(57,35){\makebox(0,0)[l]{$y$}}
\put(55,5){\makebox(0,0)[l]{$ab$}}
\put(0,20){\oval(120,30)}
\put(0,20){\makebox(0,0){$
\bcy
y=
[\![a \leq  c]\!] \, a  cd
: c   (a\to d)\\
cd=\varphi(b)\ey$}}

\put(-88,38){\makebox(0,0)[l]{$acd$}}
\put(-88,5){\makebox(0,0)[l]{$ab$}}
\put(-110,20){\oval(50,30)}
\put(-110,20){\makebox(0,0){$
\bcy
a \leq  c\\
cd=\varphi(b)\ey$}}

\put(132,38){\makebox(0,0)[l]{$cq$}}
\put(132,2){\makebox(0,0)[l]{$ab$}}
\put(112,20){\oval(54,40)}
\put(112,20){\makebox(0,0){$
\bcy
c<a\\
q=a\to d\\
cd=\varphi(b)\ey$}}

\put(145,93){\makebox(0,0)[l]{$a\varepsilon$}}
\put(145,68){\makebox(0,0)[l]{$a\varepsilon$}}
\put(140,80){\oval(20,20)}

\put(95,145){\makebox(0,0)[l]{$a\varepsilon$}}
\put(95,117){\makebox(0,0)[l]{$aij$}}
\put(80,130){\makebox(0,0){$\varepsilon=\varphi(ij)$}}
\put(80,130){\oval(45,20)}

\put(80,90){\vector(0,1){30}}

\put(78,105){\makebox(0,0)[r]{$b=ij$}}

\put(-1,53){\makebox(0,0)[r]{$p=cd$}}
\put(43,82){\makebox(0,0)[b]{$p=\varepsilon$}}
\put(115,82){\makebox(0,0)[b]{$b=\varepsilon$}}

\put(-73,23){\makebox(0,0)[b]{$a\leq c$}}
\put(73,23){\makebox(0,0)[b]{$c<a$}}

\put(-20,100){\makebox(0,0)[r]{$s_2$}}

\put(136,93){\makebox(0,0)[r]{$s_3$}}

\put(60,140){\makebox(0,0)[r]{$s^*$}}

\put(-126,40){\makebox(0,0)[r]{$s_4$}}

\put(82,40){\makebox(0,0)[l]{$s_5$}}

\put(0,63){\vector(0,-1){28}}
\put(27,80){\vector(1,0){33}}
\put(100,80){\vector(1,0){30}}

\put(60,20){\vector(1,0){25}}
\put(-60,20){\vector(-1,0){25}}

\end{picture}
\ey
\ee

One of neighborhoods
of the state 
$s^*$ in \re{sdafasdfasdf2} 
has the form
$$
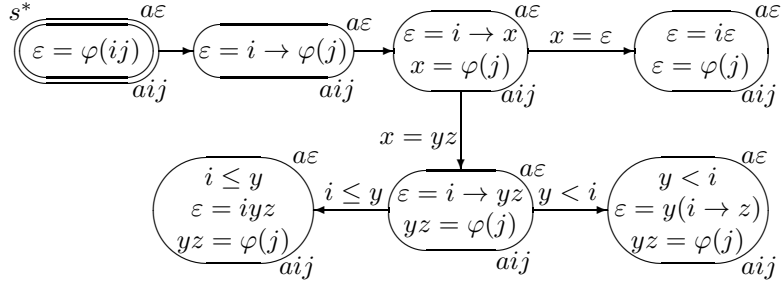
\by
\begin{picture}(250,105)

\put(20,95){\makebox(0,0)[l]{$a\varepsilon$}}
\put(17,65){\makebox(0,0)[l]{$aij$}}
\put(0,80){\oval(50,20)}
\put(0,80){\oval(54,24)}
\put(0,80){\makebox(0,0){$
\varepsilon=\varphi
(ij)$}}

\put(97,90){\makebox(0,0)[l]{$a\varepsilon$}}
\put(87,65){\makebox(0,0)[l]{$aij$}}
\put(70,80){\oval(60,20)}
\put(70,80){\makebox(0,0){$
\varepsilon=i\to \varphi
(j)$}}

\put(160,95){\makebox(0,0)[l]{$a\varepsilon$}}
\put(155,63){\makebox(0,0)[l]{$aij$}}
\put(140,80){\oval(50,30)}
\put(140,80){\makebox(0,0){$
\bcy
\varepsilon=i\to x\\
x=\varphi
(j)\ey$}}

\put(250,95){\makebox(0,0)[l]{$a\varepsilon$}}
\put(245,63){\makebox(0,0)[l]{$aij$}}
\put(230,80){\oval(50,30)}
\put(230,80){\makebox(0,0){$
\bcy
\varepsilon=i \varepsilon\\
\varepsilon=\varphi
(j)\ey$}}
\put(185,83){\makebox(0,0)[b]{$x=\varepsilon$}}

\put(-20,95){\makebox(0,0)[r]{$s^*$}}

\put(27,80){\vector(1,0){13}}
\put(100,80){\vector(1,0){15}}
\put(165,80){\vector(1,0){40}}

\put(162,35){\makebox(0,0)[l]{$a\varepsilon$}}
\put(153,0){\makebox(0,0)[l]{$aij$}}
\put(140,20){\oval(54,30)}
\put(140,20){\makebox(0,0){$\bcy
\varepsilon=i\to yz\\
yz = \varphi(j)
\ey$}}

\put(245,40){\makebox(0,0)[l]{$a\varepsilon$}}
\put(245,-2){\makebox(0,0)[l]{$aij$}}
\put(225,20){\oval(60,40)}
\put(225,20){\makebox(0,0){$\bcy
y<i\\\varepsilon=y(i\to z)\\
yz=\varphi(j)
\ey$}}

\put(77,40){\makebox(0,0)[l]{$a\varepsilon$}}
\put(72,-2){\makebox(0,0)[l]{$aij$}}
\put(55,20){\oval(60,40)}
\put(55,20){\makebox(0,0){$\bcy
i\leq y\\
\varepsilon=iyz
\\yz=\varphi(j)
\ey$}}

\put(139,46){\makebox(0,0)[r]{$x=yz$}}

\put(140,65){\vector(0,-1){30}}
\put(167,20){\vector(1,0){28}}
\put(113,20){\vector(-1,0){28}}

\put(100,22){\makebox(0,0)[b]{$i\leq y$}}

\put(180,22){\makebox(0,0)[b]{$ y<i$}}

\end{picture}
\ey
$$

All leaves of the last neighborhood are contradictory, since
among the conjunctive terms 
occurred in their conditions, there are equalities of the form
$\varepsilon=uv$, 
 which are equal to the term 
 $\bot$.
 Therefore, as it was said above,
$s^*$ is contradictory.

Bringing together
neighborhoods 
 \re{sdafasdfasdf1} and
\re{sdafasdfasdf2},
in view of the foregoing, we conclude that one of the neighborhoods of state  
$s_0\eam
\{y=\varphi(x)\}^y_x$
has the form
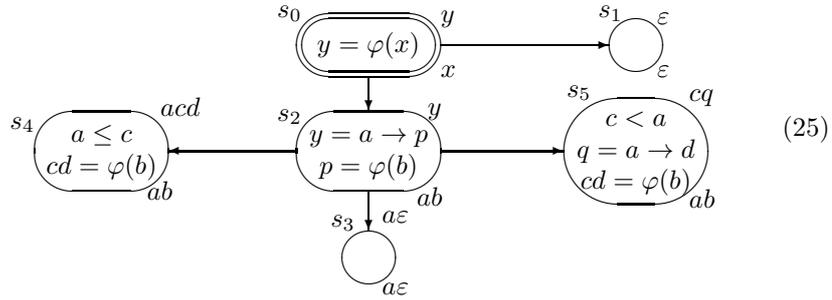
\be{fgdsgdfsgegw33erg5e334r}
\by
\begin{picture}(0,100)

\put(-25,92){\makebox(0,0)[r]{$s_0$}}
\put(27,90){\makebox(0,0)[l]{$y$}}
\put(27,70){\makebox(0,0)[l]{$x$}}
\put(0,80){\oval(50,20)}
\put(0,80){\oval(54,24)}
\put(0,80){\makebox(0,0){$
y=\varphi
(x)$}}

\put(95,92){\makebox(0,0)[r]{$s_1$}}
\put(108,90){\makebox(0,0)[l]{$\varepsilon$}}
\put(108,70){\makebox(0,0)[l]{$\varepsilon$}}
\put(100,80){\oval(20,20)}

\put(-25,53){\makebox(0,0)[r]{$s_2$}}
\put(22,55){\makebox(0,0)[l]{$y$}}
\put(18,23){\makebox(0,0)[l]{$ab$}}
\put(0,40){\oval(54,30)}
\put(0,40){\makebox(0,0){$\bcy
y=a\to p\\
p = \varphi(b)
\ey$}}

\put(-125,50){\makebox(0,0)[r]{$s_4$}}
\put(-78,57){\makebox(0,0)[l]{$acd$}}
\put(-83,25){\makebox(0,0)[l]{$ab$}}
\put(-100,40){\oval(50,30)}
\put(-100,40){\makebox(0,0){$\bcy
a\leq c\\cd=\varphi(b)
\ey$}}

\put(-5,13){\makebox(0,0)[r]{$s_3$}}
\put(5,15){\makebox(0,0)[l]{$a\varepsilon
$}}
\put(5,-12){\makebox(0,0)[l]{$a\varepsilon
$}}
\put(0,0){\oval(20,20)}

\put(74,62){\makebox(0,0)[l]{$s_5$}}
\put(120,60){\makebox(0,0)[l]{$cq$}}
\put(120,22){\makebox(0,0)[l]{$ab$}}
\put(100,40){\oval(54,40)}
\put(100,40){\makebox(0,0){$\bcy
c<a\\q=a\to d\\
cd=\varphi(b)
\ey$}}

\put(0,68){\vector(0,-1){13}}
\put(0,25){\vector(0,-1){15}}
\put(27,80){\vector(1,0){63}}
\put(-27,40){\vector(-1,0){48}}
\put(27,40){\vector(1,0){46}}

\end{picture}
\ey
\ee\\

Another example of 
a neighborhood is related to the state $s_5$.
One of neighborhoods
of this state has the form
$$\by
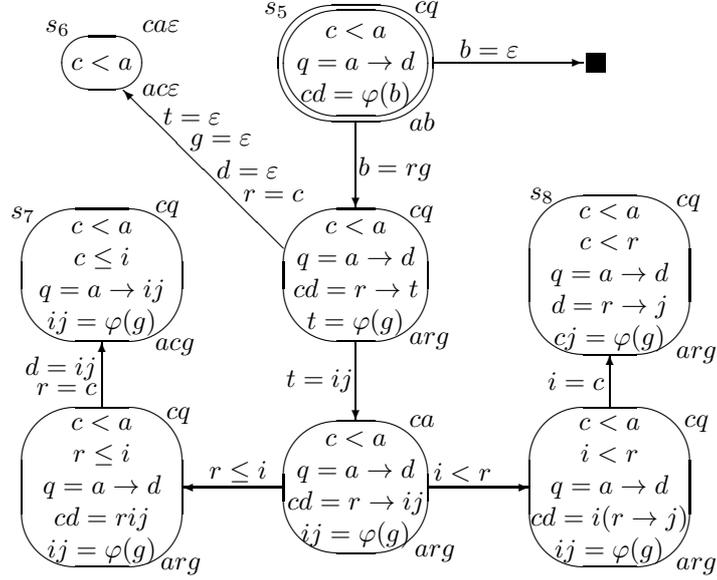
\begin{picture}(0,210)

\put(22,200){\makebox(0,0)[l]{$cq$}}
\put(20,158){\makebox(0,0)[l]{$ab$}}
\put(0,180){\oval(54,40)}
\put(0,180){\oval(58,44)}
\put(0,180){\makebox(0,0){$\bcy
c<a\\
q=a\to d\\
cd=\varphi(b)\ey$}}

\put(90,180){\makebox(0,0){$\blackbox$}}

\put(-25,200){\makebox(0,0)[r]{$s_5$}}
\put(-107,193){\makebox(0,0)[r]{$s_6$}}
\put(-120,123){\makebox(0,0)[r]{$s_7$}}
\put(65,130){\makebox(0,0)[l]{$s_8$}}

\put(-80,194){\makebox(0,0)[l]{$ca\varepsilon$}}
\put(-80,170){\makebox(0,0)[l]{$ac\varepsilon$}}
\put(-95,180){\makebox(0,0)[c]{$c<a$}}
\put(-95,180){\oval(30,20)}

\put(20,123){\makebox(0,0)[l]{$cq$}}
\put(20,75){\makebox(0,0)[l]{$arg$}}
\put(0,100){\oval(54,50)}
\put(0,100){\makebox(0,0){$\bcy
c<a\\
q=a\to d\\
cd=r\to t\\
t=\varphi(g)
\ey$}}

\put(120,127){\makebox(0,0)[l]{$cq$}}
\put(120,70){\makebox(0,0)[l]{$arg$}}
\put(95,100){\oval(60,60)}
\put(95,100){\makebox(0,0){$\bcy
c<a\\ c<r\\q=a\to d\\d=r\to j\\cj=\varphi(g)
\ey$}}

\put(20,45){\makebox(0,0)[l]{$ca$}}
\put(22,-5){\makebox(0,0)[l]{$arg$}}
\put(0,20){\oval(54,50)}
\put(0,20){\makebox(0,0){$\bcy
c<a\\q=a\to d\\cd=r\to ij\\ij=\varphi(g)
\ey$}}

\put(-71,47){\makebox(0,0)[l]{$cq$}}
\put(-73,-10){\makebox(0,0)[l]{$arg$}}
\put(-95,20){\oval(60,60)}
\put(-95,20){\makebox(0,0){$\bcy
c<a\\r\leq i\\q=a\to d\\cd=rij\\ij=\varphi(g)
\ey$}}

\put(123,44){\makebox(0,0)[l]{$cq$}}
\put(120,-10){\makebox(0,0)[l]{$arg$}}
\put(95,20){\oval(60,60)}
\put(95,20){\makebox(0,0){$\bcy
c<a\\ i<r\\q=a\to d\\cd=i(r\to j)\\ij=\varphi(g)
\ey$}}

\put(-75,125){\makebox(0,0)[l]{$cq$}}
\put(-75,73){\makebox(0,0)[l]{$acg$}}
\put(-95,100){\oval(60,50)}
\put(-95,100){\makebox(0,0){$\bcy
c<a\\c\leq i\\q=a\to ij\\ij=\varphi(g)
\ey$}}

\put(1,140){\makebox(0,0)[l]{$b=rg$}}
\put(-1,60){\makebox(0,0)[r]{$t=ij$}}
\put(-44,22){\makebox(0,0)[b]{$r\leq i$}}
\put(40,22){\makebox(0,0)[b]{$i<r$}}
\put(50,182){\makebox(0,0)[b]{$b=\varepsilon$}}
\put(-72,160){\makebox(0,0)[l]{$
t=\varepsilon$
}}
\put(-62,150){\makebox(0,0)[l]{$
g=\varepsilon$
}}
\put(-52,140){\makebox(0,0)[l]{$
d=\varepsilon$
}}
\put(-42,130){\makebox(0,0)[l]{$
r=c$
}}
\put(93,60){\makebox(0,0)[r]{$i=c$}}

\put(-97,65){\makebox(0,0)[r]{$d=ij$}}
\put(-97,57){\makebox(0,0)[r]{$r=c$}}

\put(0,158){\vector(0,-1){33}}
\put(0,75){\vector(0,-1){30}}
\put(29,180){\vector(1,0){56}}
\put(95,50){\vector(0,1){20}}
\put(-95,50){\vector(0,1){25}}
\put(-27,110){\vector(-1,1){60}}
\put(27,20){\vector(1,0){38}}
\put(-27,20){\vector(-1,0){38}}
\end{picture}
\ey
$$\\

This neighborhood can be reduced, 
and we can get
 the neighborhood
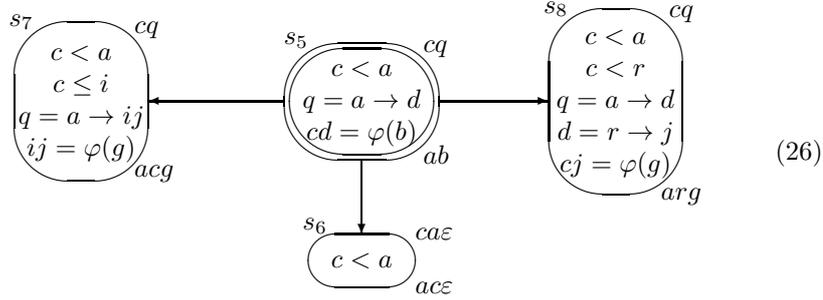
\be{sfdgdfsgsdfhgghfghfhtsh44}
\by
\begin{picture}(0,105)

\put(23,90){\makebox(0,0)[l]{$cq$}}
\put(23,50){\makebox(0,0)[l]{$ab$}}
\put(0,70){\oval(54,40)}
\put(0,70){\oval(58,44)}
\put(0,70){\makebox(0,0){$
\bcy
c<a\\
q=a\to d\\
cd=\varphi(b)\ey$}}

\put(-20,93){\makebox(0,0)[r]{$s_5$}}

\put(-132,100){\makebox(0,0)[l]{$s_7$}}
\put(68,105){\makebox(0,0)[l]{$s_8$}}

\put(-22,23){\makebox(0,0)[l]{$s_6$}}

\put(115,103){\makebox(0,0)[l]{$cq$}}
\put(112,34){\makebox(0,0)[l]{$arg$}}
\put(95,70){\oval(50,70)}
\put(95,70){\makebox(0,0){$\bcy
c<a\\ c<r\\q=a\to d\\d=r\to j\\cj=\varphi(g)
\ey$}}

\put(-85,97){\makebox(0,0)[l]{$cq$}}
\put(-85,43){\makebox(0,0)[l]{$acg$}}
\put(-105,70){\oval(50,60)}
\put(-105,70){\makebox(0,0){$\bcy
c<a\\c\leq i\\q=a\to ij\\ij=\varphi(g)
\ey$}}

\put(0,48){\vector(0,-1){28}}
\put(29,70){\vector(1,0){41}}
\put(-29,70){\vector(-1,0){51}}

\put(20,20){\makebox(0,0)[l]{$ca\varepsilon
$}}
\put(20,0){\makebox(0,0)[l]{$ac\varepsilon
$}}
\put(0,10){\makebox(0,0){$c<a$}}
\put(0,10){\oval(40,20)}

\end{picture}
\ey
\ee

\subsubsection{Examples 
of 
neighborhoods for string ordering checking program}

Other examples of neighborhoods
of states are related to the FP 
\re{sdfgfdsgsdfgdsfgsrrr}
of  string ordering checking.
We rewrite this FP using a shorter notation for the function variable occurred  in it:
\be{afgdfssdfgdsgw4g455w4}
o(x) 
=[\![x = \varepsilon]\!] 1
:\Big( [\![x_t = \varepsilon]\!] 1
: [\![x_h \leq (x_t)_h]\!] o(x_t): 0\Big)
\ee

One of neighborhoods of the state 
$\sigma_0\eam
\{z=o(h)\}^z_h$ of this
FP 
has the form 
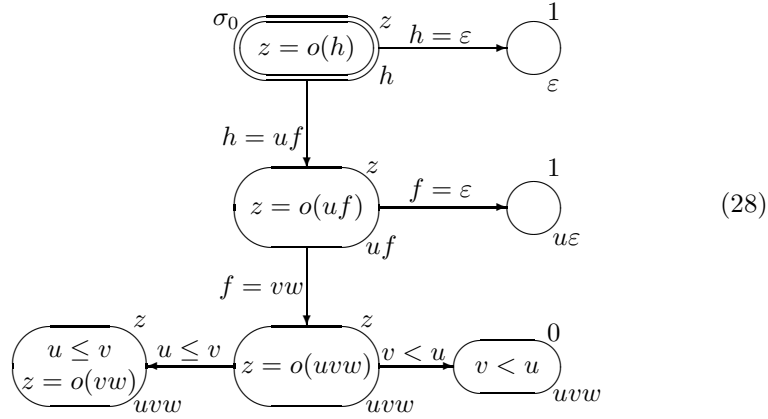
\be{fsdgdfshdghdgh5e6h56h5h6e5}
\by
\begin{picture}(0,165)

\put(27,150){\makebox(0,0)[l]{$z$}}
\put(27,130){\makebox(0,0)[l]{$h$}}
\put(0,140){\oval(50,20)}
\put(0,140){\oval(54,24)}
\put(0,140){\makebox(0,0){$
z=o
(h)$}}

\put(-25,150){\makebox(0,0)[r]{$\sigma_0$}}

\put(90,154){\makebox(0,0)[l]{$1$}}
\put(90,126){\makebox(0,0)[l]{$\varepsilon$}}
\put(85,140){\oval(20,20)}

\put(22,95){\makebox(0,0)[l]{$z$}}
\put(22,65){\makebox(0,0)[l]{$uf$}}
\put(0,80){\oval(54,30)}
\put(0,80){\makebox(0,0){$\bcy
z=o(uf)
\ey$}}

\put(90,95){\makebox(0,0)[l]{$1
$}}
\put(92,68){\makebox(0,0)[l]{$u\varepsilon$}}
\put(85,80){\oval(20,20)}

\put(20,37){\makebox(0,0)[l]{$z$}}
\put(22,5){\makebox(0,0)[l]{$uvw$}}
\put(0,20){\oval(54,30)}
\put(0,20){\makebox(0,0){$\bcy
z=o(uvw)
\ey$}}

\put(90,33){\makebox(0,0)[l]{$0$}}
\put(92,9){\makebox(0,0)[l]{$uvw$}}
\put(75,20){\oval(40,20)}
\put(75,20){\makebox(0,0){$\bcy
v<u
\ey$}}

\put(-65,37){\makebox(0,0)[l]{$z$}}
\put(-65,5){\makebox(0,0)[l]{$uvw$}}
\put(-85,20){\oval(50,30)}
\put(-85,20){\makebox(0,0){$\bcy
u\leq v\\
z=o(vw)
\ey$}}

\put(-1,106){\makebox(0,0)[r]{$h=uf$}}
\put(-1,50){\makebox(0,0)[r]{$f=vw$}}
\put(-44,22){\makebox(0,0)[b]{$u\leq v$}}
\put(40,22){\makebox(0,0)[b]{$v<u$}}
\put(50,142){\makebox(0,0)[b]{$h=\varepsilon$}}
\put(50,82){\makebox(0,0)[b]{$f=\varepsilon$}}

\put(0,128){\vector(0,-1){33}}
\put(0,65){\vector(0,-1){30}}
\put(27,140){\vector(1,0){48}}
\put(27,80){\vector(1,0){48}}
\put(27,20){\vector(1,0){28}}
\put(-27,20){\vector(-1,0){33}}

\end{picture}
\ey
\ee

Using the definition of the concept of a neighborhood of a state of a FP, 
\re{fsdgdfshdghdgh5e6h56h5h6e5}
can be transformed to
the neighborhood
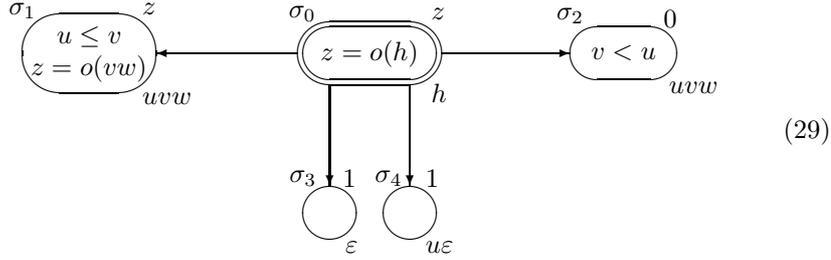
\be{asdfadsgfdsgdfsgsdfgdfsg}
\by
\begin{picture}(0,105)

\put(23,95){\makebox(0,0)[l]{$z$}}
\put(23,65){\makebox(0,0)[l]{$h$}}
\put(0,80){\oval(50,20)}
\put(0,80){\oval(54,24)}
\put(0,80){\makebox(0,0){$
z=o
(h)$}}

\put(-20,95){\makebox(0,0)[r]{$\sigma_0$}}
\put(-135,96){\makebox(0,0)[l]{$\sigma_1$}}
\put(70,94){\makebox(0,0)[l]{$\sigma_2$}}
\put(-30,33){\makebox(0,0)[l]{$\sigma_3$}}
\put(2,33){\makebox(0,0)[l]{$\sigma_4$}}

\put(110,93){\makebox(0,0)[l]{$0$}}
\put(112,67){\makebox(0,0)[l]{$uvw$}}
\put(95,80){\oval(40,20)}
\put(95,80){\makebox(0,0){$\bcy
v<u
\ey$}}

\put(-85,97){\makebox(0,0)[l]{$z$}}
\put(-85,63){\makebox(0,0)[l]{$uvw$}}
\put(-105,80){\oval(50,30)}
\put(-105,80){\makebox(0,0){$\bcy
u\leq v\\
z=o(vw)
\ey$}}

\put(-15,68){\vector(0,-1){38}}
\put(15,68){\vector(0,-1){38}}
\put(27,80){\vector(1,0){48}}
\put(-27,80){\vector(-1,0){53}}

\put(21,33){\makebox(0,0)[l]{$1
$}}
\put(21,7){\makebox(0,0)[l]{$u\varepsilon
$}}
\put(15,20){\oval(20,20)}
\put(-10,33){\makebox(0,0)[l]{$1
$}}
\put(-9,7){\makebox(0,0)[l]{$ \varepsilon
$}}
\put(-15,20){\oval(20,20)}

\end{picture}
\ey
\ee

\section{Embeddings of states
of functional programs}

\subsection{
Explicit, conditional and justified
embeddings}

Let
$s,s'$ be states from ${\cal S}_\Sigma$.
\bi\i
An {\bf explicit embedding}
 $s$ in $s'$ is a notation of the form
 $$
\theta:s\hra{}s',$$
 where $\theta$ is a clarification, and 
$\beta_s= \beta_{s'}[\theta]\wedge
\beta$, and $\Phi_\beta=\emptyset$. 

\i A {\bf conditional embedding} $s$ in $s'$ is a notation of the form
\be{sdfdsgdfsgdsfgsdhgfhgd}
\c{\eta:
r\hra{} r'\\
 u[\theta]\hookrightarrow 
 u'[\theta']
}:
s\hookrightarrow s',\ee
 where 
$\eta: r\hra{} r'$ is an
explicit embedding, 
$u$,
$u'$
$\in {\cal S}_\Sigma$,
$\theta$ and $\theta'$ are clarifications, 
and
$$
\beta_{s}= \beta_{u[\theta]}
\wedge 
\beta_{r},\quad
\beta_{s'}= \beta_{u'[\theta']}
\wedge
\beta_{r'}.$$

A  {\bf premise} of the 
conditional embedding \re{sdfdsgdfsgdsfgsdhgfhgd}
is a notation 
$u\hookrightarrow
 u'$, where $u$ and $u'$
are correspongins states occurred in 
 \re{sdfdsgdfsgdsfgsdhgfhgd}.
\i A {\bf justified} embedding
of $s$ in $s'$ is a notation of the form
 \be{dfsgdsfgds456345634}
 s\hra{!} s'\ee
if 
$\exists\,U\in {\cal U}_s$, 
$\exists\,U'\in {\cal U}_{s'}$:
for each 
non-terminal leaf $r\in U$
\bi
\i either there is an 
explicit embedding
$r$ in some $r'\in U'$,\i
or there is a conditional embedding
$r$ in some $r'\in U'$, 
and its premise has the form
$s\hookrightarrow s'$, 
 where $s$ and $s'$ are states from
\re{dfsgdsfgds456345634}. 
\ei\i
A state $s$ is said to be
{\bf embedded} in $s'$,
if there is \bi\i either 
explicit embedding
$s$ in $s'$, \i or
conditional embedding 
$s$ in $s'$ with a justified 
premise.\ei

The notation 
$s\subseteq s'$ 
means that 
$s$ is embedded in $s'$.
\ei

Note that each justified embedding can be considered as a conditional embedding with a justified premise (an ``explicit embedding '' component in this conditional embedding is trivial).

\subsection{Examples of embeddings of states}

\subsubsection{Examples of explicit embeddings of states}
\bn
\i
For the states
$s_4=\c{a\leq c\\cd=\varphi(b)}^{acd}_{ab}$ and $s_0=
\{y=\varphi(x)\}^y_x$,  
occurred in  neighborhood
\re{fgdsgdfsgegw33erg5e334r}, there is an explicit embedding
\be{sdfdsfafer44456dgfdfsg}
[cd/y,b/x]:s_4\hookrightarrow s_0.\ee
\i For the states 
$s_7=\c{c<a,\;c\leq i\\q=a\to ij\\ij=\varphi(g)}^{cq}_{acg}$
and
$s_2=
\c{y=a\to p\\p=\varphi(b)}^y_{ab}$, 
occurred in  neighborhoods
\re{sfdgdfsgsdfhgghfghfhtsh44} and
\re{sdafasdfasdf1} respectively, 
there is an explicit embedding
\be{sdfgdsgerwere8rw334}
[q/y,ij/p,g/b]:
s_7\hookrightarrow s_2.\ee
\i  For the states
$\sigma_1=\c{r\leq v
\\
z=o(vw)}^{z}_{rvw}
$ and
$\sigma_0 = \{z=o(h)\}^z_h
$, occurred in neighborhood 
\re{asdfadsgfdsgdfsgsdfgdfsg},
there is an explicit  embedding
\be{dsafasferf4455553}
[vw/h]:\sigma_1
\hookrightarrow 
\sigma_0.\ee
\en
\subsubsection{An example
of conditional 
embedding}

 For the states
$s_8=\c{
c<a,\; c<r\\q=a\to d\\d=r\to j\\cj=\varphi(g)}^{cq}_{arg}$ and
$s_2=
\c{y=a\to p\\p=\varphi(b)}^y_{ab},$
 occurred in neighborhoods
\re{sfdgdfsgsdfhgghfghfhtsh44} and
\re{sdafasdfasdf1} respectively,
there is a 
conditional embedding
with the premise 
$s_5\hookrightarrow
s_0$:
\be{sdfdsgdfsgdsfgsdhgfh4453gd}
\c{
[q/y,d/p]:\c{
c<a\\q=a\to d}^{cq}_{acd}\hookrightarrow 
 \{y=a\to p\}^{y}_{ap}\\
s_5[\theta]=
\c{
c<r\\d=r\to j\\cj=\varphi(g)}^{cd}_{rg}
\hookrightarrow
 s_0[\theta']=
 \{p=\varphi(b)\}^{p}_{b}
}:
s_8\hookrightarrow s_2,\ee
 where $\theta=[r/a,d/q,j/d,g/b]$,
$\theta'=[p/y,b/x]$.

\subsubsection{An example of 
a justified embedding}
 \label{dfgdsfhgfghfghdfgh}

An example of a justified 
embedding is
 $s_5\hra{!} s_0$,
 where $s_5$ and $s_0$ are states from 
\re{fgdsgdfsgegw33erg5e334r}.
In this case $U=
\re{sfdgdfsgsdfhgghfghfhtsh44}$
and
$U'=
\re{sdafasdfasdf1}$.

In the neighborhood \re{sfdgdfsgsdfhgghfghfhtsh44}
\bi
\i  state $s_6$ is terminal,
\i there is explicit embedding
\re{sdfgdsgerwere8rw334}
of state $s_7$ in state
$s_2$, and 
\i there is a
conditional embedding
\re{sdfdsgdfsgdsfgsdhgfh4453gd}
of state $s_8$ in state
$s_2$
with the premise $s_5\hookrightarrow s_0$.
\ei

\section{State diagrams}
\label{sadfsadfasd3334}

\subsection{A concept of a state diagram}

Let $\Sigma$ be a FP.
A {\bf state diagram (SD)} of $\Sigma$ is a triple
\be{sdfgdsgsdfd45366}D=(U,N, I),\ee whose components
have the following meaning:\bi
\i $U$ is a neighborhood
of the initial state
$s^0_\Sigma$, 
\i $N$ is a set of all 
non-terminal leaves of $U$, and
\i  $I$ is a set of pairs of the form
$(s,s')$,
 where $s\in N$, $s'$ is an
 ancestor of $s$,
 $s\subseteq s'$, and $\forall\,
s\in N\;\;\exists\,s':(s,s')\in I$.
\ei

In the graphic representation 
of SD
\re{sdfgdsgsdfd45366}
we will denote pairs
from $I$ by labelled 
 arrows on the neighborhood
$U$: a pair
$(s,s')\in I$ 
 will be represented by an arrow starting from  $s$,
ending to $s'$ and labelled by
$\subseteq$.

\subsection{Examples of state diagrams}

\subsubsection{State diagram for a sorting program}

SD for FP  
\re{fdsgdsgds33r} is based on neighborhood
\re{fgdsgdfsgegw33erg5e334r}
and has the form

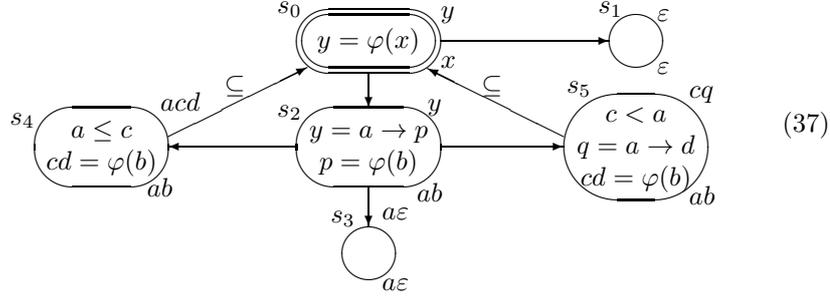
\be{fgdsgdfsgegw33erg5e332334r}
\by
\begin{picture}(0,100)

\put(-25,92){\makebox(0,0)[r]{$s_0$}}
\put(27,90){\makebox(0,0)[l]{$y$}}
\put(27,72){\makebox(0,0)[l]{$x$}}
\put(0,80){\oval(50,20)}
\put(0,80){\oval(54,24)}
\put(0,80){\makebox(0,0){$
y=\varphi
(x)$}}

\put(95,92){\makebox(0,0)[r]{$s_1$}}
\put(108,90){\makebox(0,0)[l]{$\varepsilon$}}
\put(108,70){\makebox(0,0)[l]{$\varepsilon$}}
\put(100,80){\oval(20,20)}

\put(-25,53){\makebox(0,0)[r]{$s_2$}}
\put(22,55){\makebox(0,0)[l]{$y$}}
\put(18,23){\makebox(0,0)[l]{$ab$}}
\put(0,40){\oval(54,30)}
\put(0,40){\makebox(0,0){$\bcy
y=a\to p\\
p = \varphi(b)
\ey$}}

\put(-125,50){\makebox(0,0)[r]{$s_4$}}
\put(-78,57){\makebox(0,0)[l]{$acd$}}
\put(-83,25){\makebox(0,0)[l]{$ab$}}
\put(-100,40){\oval(50,30)}
\put(-100,40){\makebox(0,0){$\bcy
a\leq c\\cd=\varphi(b)
\ey$}}

\put(-5,13){\makebox(0,0)[r]{$s_3$}}
\put(5,15){\makebox(0,0)[l]{$a\varepsilon
$}}
\put(5,-12){\makebox(0,0)[l]{$a\varepsilon
$}}
\put(0,0){\oval(20,20)}

\put(74,62){\makebox(0,0)[l]{$s_5$}}
\put(120,60){\makebox(0,0)[l]{$cq$}}
\put(120,22){\makebox(0,0)[l]{$ab$}}
\put(100,40){\oval(54,40)}
\put(100,40){\makebox(0,0){$\bcy
c<a\\q=a\to d\\
cd=\varphi(b)
\ey$}}

\put(0,68){\vector(0,-1){13}}
\put(0,25){\vector(0,-1){15}}
\put(27,80){\vector(1,0){63}}
\put(-27,40){\vector(-1,0){48}}
\put(27,40){\vector(1,0){46}}

\put(-75,44){\vector(2,1){52}}
\put(73,44){\vector(-2,1){51}}

\put(-54,63){\makebox(0,0)[l]{$\subseteq$}}
\put(50,63){\makebox(0,0)[r]{$\subseteq$}}

\end{picture}
\ey
\ee\\

In this SD, the edges labeled  by
$\subseteq$
correspond to explicit embedding \re{sdfdsfafer44456dgfdfsg}
and justified embedding
$s_5\hra{!} s_0$,
considered in
\ref{dfgdsfhgfghfghdfgh}.

\subsubsection{State diagram for the program
of string ordering checking} 

SD for FP  
\re{afgdfssdfgdsgw4g455w4}
is built on the base of neighborhood \re{asdfadsgfdsgdfsgsdfgdfsg}
and has the form
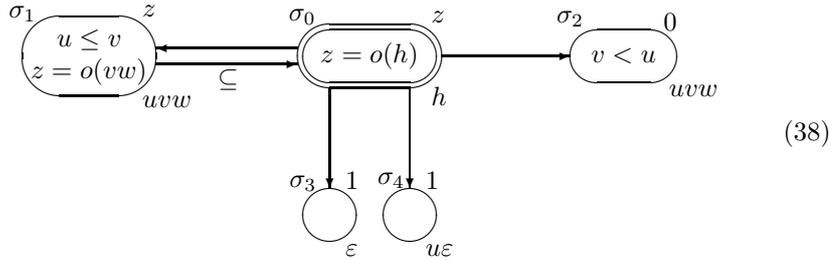
\be{asdfadsgfdsgdfsg22sdfgdfsg}
\by
\begin{picture}(0,105)

\put(23,95){\makebox(0,0)[l]{$z$}}
\put(23,65){\makebox(0,0)[l]{$h$}}
\put(0,80){\oval(50,20)}
\put(0,80){\oval(54,24)}
\put(0,80){\makebox(0,0){$
z=o
(h)$}}

\put(-20,95){\makebox(0,0)[r]{$\sigma_0$}}
\put(-135,96){\makebox(0,0)[l]{$\sigma_1$}}
\put(70,94){\makebox(0,0)[l]{$\sigma_2$}}
\put(-30,32){\makebox(0,0)[l]{$\sigma_3$}}
\put(3,33){\makebox(0,0)[l]{$\sigma_4$}}

\put(110,93){\makebox(0,0)[l]{$0$}}
\put(112,67){\makebox(0,0)[l]{$uvw$}}
\put(95,80){\oval(40,20)}
\put(95,80){\makebox(0,0){$\bcy
v<u
\ey$}}

\put(-85,97){\makebox(0,0)[l]{$z$}}
\put(-85,63){\makebox(0,0)[l]{$uvw$}}
\put(-105,80){\oval(50,30)}
\put(-105,80){\makebox(0,0){$\bcy
u\leq v\\
z=o(vw)
\ey$}}

\put(-53,75){\makebox(0,0)[t]{$\subseteq$}}

\put(-15,68){\vector(0,-1){38}}
\put(15,68){\vector(0,-1){38}}
\put(27,80){\vector(1,0){48}}
\put(-27,83){\vector(-1,0){53}}
\put(-80,77){\vector(1,0){53}}

\put(21,33){\makebox(0,0)[l]{$1
$}}
\put(21,7){\makebox(0,0)[l]{$u\varepsilon
$}}
\put(15,20){\oval(20,20)}
\put(-9,33){\makebox(0,0)[l]{$1
$}}
\put(-9,7){\makebox(0,0)[l]{$\varepsilon
$}}
\put(-15,20){\oval(20,20)}

\end{picture}
\ey
\ee

In this SD, an edge labeled by 
$\subseteq$ corresponds to explicit embedding
\re{dsafasferf4455553}.

 \section{Verification of functional programs based on the concept of a state diagram}

\subsection{Composition of functional programs}

\subsubsection{The concept of a composition of functional programs}
\label{fdgdsgdfsssdf}

Let $\Sigma$ and
$\Sigma'$ be FPs, and 
main terms in
$\Sigma$ and
$\Sigma'$
have the form  $\varphi(\bar x)$ and
$\varphi'(u)$ respectively, 
where
 $\tau(\varphi(\bar x))=\tau(u)$, and
 $X\Phi_\Sigma\cap
X\Phi_{\Sigma'}=\emptyset$.

In this case, it can be defined
a FP $\Sigma'(\Sigma)$, 
called a 
{\bf composition} of  FPs  
$\Sigma$ and $\Sigma'$,
and is a set of equalities,
\bi\i
the first of which
has the form
$
\psi(\bar x)=
\varphi'(\varphi(\bar x))$,
 where  $\psi$ 
is a  fresh functional
a variable of the appropriate type,
and \i other equalities
 are all equalities,
 occurred in  
$\Sigma$ and $\Sigma'$.\ei

It is easy to see that
$\forall\,\bar d\in {\cal D}_{\bar x}\quad
f_{\Sigma'(\Sigma)}(\bar d)=
f_{\Sigma'}(f_{\Sigma}(\bar d))$.
 
 \subsubsection{Neighborhoods
of an initial state of a composition of 
functional programs}
\label{sdgsdfgsdfhghsdfr44}

Let \bi\i $\Sigma$ and
$\Sigma'$ be FPs 
satisfying
conditions 
at the beginning of 
item 
\ref{fdgdsgdfsssdf},
and
\i
$U\in {\cal U}_{s^0_{\Sigma}}$, 
$U'\in {\cal U}_{s^0_{\Sigma'}}$
be neighborhoods 
such that
$$\forall\,s\in U,\;
\forall\,s'\in U'\quad
X_{s}\cap X_{s'}=\emptyset.$$
\ei

$\forall\,s\in U$, $\forall\,s'\in U'$
it will be denoted by
$ss'$ a state
FP $\Sigma'(\Sigma)$, 
defined as follows:
let $s=\{\beta\}^{y}_{\bar x}$,
$s'=\{\beta'\}^{z}_{y'}$,
then
$$ss'\eam \{\beta\wedge \beta'\wedge
(y=y')\}^z_{\bar x}.$$

It is easy to see that if
$s^0_{\Sigma}=\{y=\varphi(\bar x)\}^y_{\bar x}$ and
$s^0_{\Sigma'}=\{z=\varphi'(y')\}^z_{y'}$, 
then
$$s^0_{\Sigma'(\Sigma)}=
\{z=\psi(\bar x)\}^z_{\bar x}=
\{z=\varphi'(\varphi(\bar x))\}^z_{\bar x}=\c{z=\varphi'(y)\\
y=\varphi(\bar x)}^z_{\bar x}=
s^0_{\Sigma}s^0_{\Sigma'}.$$

Let $UU'$ be a tree, \bi\i
nodes of which have labels of the 
form
$ss'$, where $s\in U$, $s'\in U'$,
and
\i which is defined by an
a non-deterministic algorithm for its construction.
\ei
The algorithm of construction
of the tree $UU'$
consists of several stages.
A tree built at each of these stages is denoted by the same notation $UU'$.
\bi
\i At the first stage,
$UU'$  is defined as a tree from one node that has the label
$s^0_{\Sigma}s^0_{\Sigma'}$.
\i Each subsequent step is that if the tree $UU'$  constructed so far contains a
leaf $v$ labeled by
$ss'$, where either $s$ 
 is not a leaf in $U$, or
$s'$  is not a leaf in  $U'$ 
 then one of the following two operations is performed:
\bi
\i if $s$  is not a leaf in  
$U$, and 
the list of its followers is of the form
$s_1,\ldots, s_n$,
then the followers of the node $v$
with the labels 
$s_1s'$, $\ldots$, $s_ns'$,
are added to the constructed tree
$UU'$,
\i if $s'$ is not a leaf in $U'$, then 
 then instead of the previous operation  a similar operation can be performed for followers of  $s'$.
\ei
\ei

\refstepcounter{theorem}
{\bf Theorem\arabic{theorem}\label{th1}}

The above  tree $UU'$
is a neighborhood
of an initial state of FP $\Sigma'(\Sigma)$.
$\blackbox$
 
 \subsubsection{
A state diagram of a composition of 
functional programs}

\refstepcounter{theorem}
{\bf Theorem\arabic{theorem}\label{th2}}

Let FPs
$\Sigma$ and $\Sigma'$
have SDs, and the  composition
$\Sigma'(\Sigma)$ is defined.

Then FP
$\Sigma'(\Sigma)$
also has SD. $\blackbox$

\subsection{The problem  of verification of functional programs}

{\bf The problem  of verification} 
of a FP 
$\Sigma$ is
in constructing the proof of the statement  that FP
 $\Sigma$ 
  satisfies
  property expressed
 by  some formal
  specification $Spec$.
  
  Below we shall  use the
following agreement:
the notation of the form $f=1$, 
 where $f$ is a function, denotes
the following statement:$$\mbox{
the function
$f$ has a value 1 
 on all its arguments.}$$

In some cases
\bi\i  formal
  specification
  $Spec$ is expressed by another FP 
$\Sigma'$, and 
\i  correctness
$\Sigma$ with respect to 
$Spec$ is represented by the statement
\be{dsfasdgdfgdsfgds}
f_{\Sigma'(\Sigma)}=1.\ee 
\ei

For example, one of the 
correctness properties of 
sorting FP in section
\ref{zsdfgdfsgdsfgdfsgsdffgds1}
is expressed by a statement of the form 
\re{dsfasdgdfgdsfgds}
(namely, by 
the statement 
\re{4.21}).\\

\refstepcounter{theorem}
{\bf Theorem\arabic{theorem}\label{th3}}

Let FP $\Sigma$ has
a SD, in which for each terminal state $s$
the term
$y_s$  is a constant  $1$.
Then $f_\Sigma=1$.
$\blackbox $\\

The above theorems are the theoretical basis of the FP verification method based on the construction of SD
\bi\i
for the analyzed FP $\Sigma$, 
and \i for FP $\Sigma'$, 
 representing the property being checked.\ei
If these FPs have SDs, then, according to the theorem \ref{th2}, 
$\Sigma'(\Sigma)$ also has a SD.
If this SD has the property described in theorem 
\ref{th3}, then 
\re{dsfasdgdfgdsfgds} holds.

The following section provides an example of the application of this method.

\subsection {An example of verification of a sorting functional program using a state diagram}

In this section, we illustrate the verification method described above with an example of the proof of the statement \re{4.21} for FP defined in
section
\ref{zsdfgdfsgdsfgdfsgsdffgds1}.

To prove equality
 \re{dsfasdgdfgdsfgds},
 where $\Sigma =$
\re{fdsgdsgds33r}
 and $\Sigma' =$
\re{afgdfssdfgdsgw4g455w4},
we construct a
neighborhood of the initial state
$s^0_{\Sigma'(\Sigma)}$
FP $\Sigma'(\Sigma)$
as a 
neighborhood if the form $UU'$,
according to the algorithm 
at  \re{sdgsdfgsdfhghsdfr44}
where
\bi
\i $U$ is the neighborhood 
\re{fgdsgdfsgegw33erg5e334r}
of the state $s^0_{\Sigma}$
and \i $U'$ is the 
neighborhood 
\re{asdfadsgfdsgdfsgsdfgdfsg}
of the state $s^0_{\Sigma'}$.
\ei

Some states of the resulting neighborhood will be contradictory.
After removing them, we get the following neighborhood:
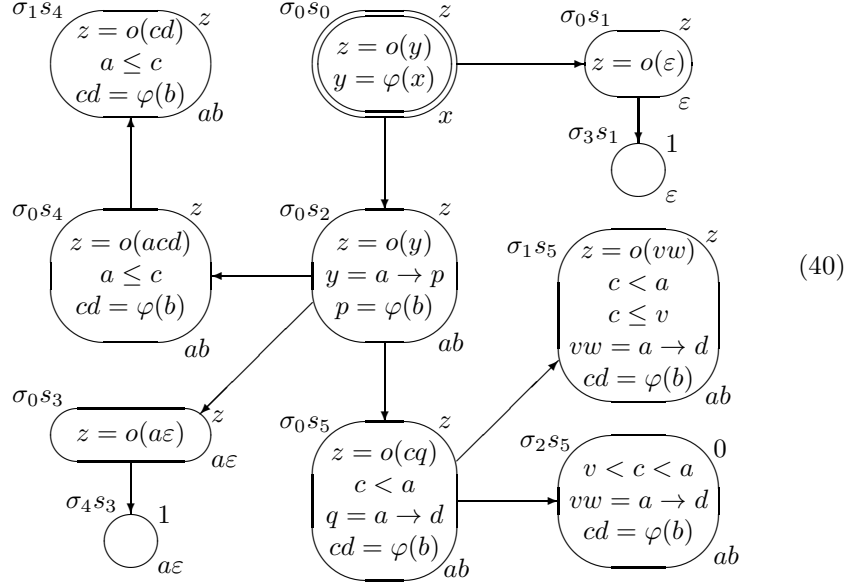
\be{dfsgdsfhdghdghdf555}
\by
\begin{picture}(0,210)

\put(105,150){\makebox(0,0)[l]{$1$}}
\put(105,130){\makebox(0,0)[l]{$\varepsilon$}}
\put(95,140){\oval(20,20)}

\put(20,200){\makebox(0,0)[l]{$z$}}
\put(20,160){\makebox(0,0)[l]{$x$}}
\put(0,180){\oval(54,40)}
\put(0,180){\oval(50,36)}
\put(0,180){\makebox(0,0){$\bcy
z=o(y)\\
y=\varphi(x)\ey$}}

\put(95,180){\oval(40,25)}
\put(95,180){\makebox(0,0){$\bcy
z=o(\varepsilon)
\ey$}}

\put(-70,197){\makebox(0,0)[l]{$z$}}
\put(-70,162){\makebox(0,0)[l]{$ab$}}
\put(-95,180){\makebox(0,0)[c]{$\bcy
z=o(cd)\\
a\leq c\\cd=\varphi(b)
\ey$}}
\put(-95,180){\oval(60,40)}

\put(20,125){\makebox(0,0)[l]{$z$}}
\put(20,75){\makebox(0,0)[l]{$ab$}}
\put(0,100){\oval(54,50)}
\put(0,100){\makebox(0,0){$\bcy
z=o(y)\\
y=a\to p\\
p=\varphi(b)
\ey$}}

\put(120,115){\makebox(0,0)[l]{$z$}}
\put(120,55){\makebox(0,0)[l]{$ab$}}
\put(95,85){\oval(60,65)}
\put(95,85){\makebox(0,0){$\bcy
z=o(vw)\\c<a\\c\leq v\\
vw=a\to d\\
cd=\varphi(b)
\ey$}}

\put(20,45){\makebox(0,0)[l]{$z$}}
\put(22,-10){\makebox(0,0)[l]{$ab$}}
\put(0,15){\oval(54,60)}
\put(0,15){\makebox(0,0){$\bcy
z=o(cq)\\
c<a\\
q=a\to d\\
cd=\varphi(b)
\ey$}}


\put(-65,47){\makebox(0,0)[l]{$z$}}
\put(-65,30){\makebox(0,0)[l]{$a\varepsilon$}}
\put(-95,40){\oval(60,20)}
\put(-95,40){\makebox(0,0){$\bcy
z=o(a\varepsilon)\ey$}}

\put(-85,10){\makebox(0,0)[l]{$1$}}
\put(-85,-10){\makebox(0,0)[l]{$a\varepsilon$}}
\put(-95,0){\oval(20,20)}
\put(-95,0){\makebox(0,0){$\bcy
\ey$}}

\put(88,153){\makebox(0,0)[r]{$\sigma_3s_1$}}
\put(-20,200){\makebox(0,0)[r]{$\sigma_0s_0$}}
\put(85,197){\makebox(0,0)[r]{$\sigma_0s_1$}}
\put(110,195){\makebox(0,0)[l]{$z$}}
\put(110,165){\makebox(0,0)[l]{$\varepsilon$}}
\put(-120,200){\makebox(0,0)[r]{$\sigma_1s_4$}}
\put(-120,124){\makebox(0,0)[r]{$\sigma_0s_4$}}
\put(-120,54){\makebox(0,0)[r]{$\sigma_0s_3$}}
\put(-100,14){\makebox(0,0)[r]{$\sigma_4s_3$}}
\put(-20,124){\makebox(0,0)[r]{$\sigma_0s_2$}}
\put(-20,44){\makebox(0,0)[r]{$\sigma_0s_5$}}

\put(65,110){\makebox(0,0)[r]{$\sigma_1s_5$}}

\put(70,37){\makebox(0,0)[r]{$\sigma_2s_5$}}
\put(123,35){\makebox(0,0)[l]{$0$}}
\put(123,-5){\makebox(0,0)[l]{$ab$}}
\put(95,15){\oval(60,50)}
\put(95,15){\makebox(0,0){$\bcy
v<c<a\\
vw=a\to d\\
cd=\varphi(b)
\ey$}}

\put(-73,125){\makebox(0,0)[l]{$z$}}
\put(-75,73){\makebox(0,0)[l]{$ab$}}
\put(-95,100){\oval(60,50)}
\put(-95,100){\makebox(0,0){$\bcy
z=o(acd)\\
a\leq c\\cd=\varphi(b)
\ey$}}



\put(27,180){\vector(1,0){48}}

\put(95,167){\vector(0,-1){17}}


\put(0,160){\vector(0,-1){35}}
\put(0,75){\vector(0,-1){30}}
\put(-27,100){\vector(-1,0){38}}
\put(-95,125){\vector(0,1){35}}
\put(-95,30){\vector(0,-1){20}}

\put(-27,90){\vector(-1,-1){42}}
\put(27,30){\vector(1,1){38}}

\put(27,15){\vector(1,0){38}}

\end{picture}
\ey
\ee\\

Neighborhood \re{dfsgdsfhdghdghdf555}
has 5 leaves. 
It is easy to see that
\bi
\i output term of two of
these leaves
($\sigma_3s_1$ and 
$\sigma_4s_3$) is equal to 1,
\i these is an explicit embedding
of the leaf $\sigma_1s_4$ 
to the state $\sigma_0s_0$:
$$[b/x,cd/y]:\sigma_1s_4\hookrightarrow
\sigma_0s_0,$$ 
\i there is a conditional embedding
 $\sigma_1s_5$
to $\sigma_0s_0$
with a justified
premise $s_5\hookrightarrow s_0$:
$$\c{
[vw/y]:\c{
c\leq v\\z=o(vw)}^{z}_{vw}\hookrightarrow 
\{z=o(y)\}^z_y\\
s_5[vw/q]\hookrightarrow s_0[\;]
}:
\sigma_1s_5\hookrightarrow 
\sigma_0s_0.$$\ei

Let is construct a neighborhood 
of the leaf $\sigma_2s_5$.
Consider followers
of the state $\sigma_2s_5$,
corresponding to followers
$s_6,s_7,s_8$ of the state $s_5$.
\bi
\i $\sigma_2s_6=\{v<u,c<a,
uvw=ca\varepsilon\}^0_{ac\varepsilon}$, this state is contradictory.
\i $\sigma_2s_7=\c{v<c<a,c\leq i\\
vw=a\to ij\\
ij=\varphi(b)\\}^0_{acg}$.
There are two complementary transitions from this state to states,
\bi\i one of which has a conjunctive term
$v=a$ in its condition, and 
\i another state has a conjunctive term
$v=i$ in its condition.
\ei
It is easy to see that both of these states are
contradictory.
\i $\sigma_2s_8=
\c{v<c<a,c<r\\
vw=a\to d\\d=r\to j\\cj=\varphi(g)\\}^0_{arg}$.
There are two complementary transitions from this state to states
$s,s'$, 
where
\bi\i $\beta_s$
has a conjunctive term
$d=\varepsilon$,
thus
  $\beta_s$ 
  has a conjunctive term
$v=a$,
where is it easy to get
a conjunctive term
$a<c<a$ in $\beta_s$, i.e. 
$s$ is contradictory,
\i 
$\beta_{s'}$ has a conjunctive term
 $d=pq$,
 where $p,q$ are fresh variables.
There are two complementary transitions from 
$s'$ to the states
$\tilde s,\tilde s'$, 
where\bi\i
$\beta_{\tilde s}$
has a conjunctive term
$a\leq p$, whence it follows that
there is a conjunctive term $v=a$
  in $\beta_{\tilde s}$,
  where it is easy to prove that $\tilde s$ is contradictory,
and \i there is a conjunctive term 
$p<a$
in $\beta_{\tilde s'}$, and
$$\tilde s'=\c{v<c<a,c<r\\
w=a\to q\\vq=r\to j\\
cj=\varphi(g)}^0_{arg}=
\c{v<c<a,c<r\\
vq=r\to j\\
cj=\varphi(g)}^0_{arg}.$$
\ei\ei
\ei
Thus, one of neighborhoods of 
$\sigma_2s_5$
has the form
\be{sdfgdfsgds345264435}\sigma_2s_5\to \tilde s'.\ee

There is an explicit embedding
$$[q/w,r/a,j/d,g/b]:
\tilde s'\hookrightarrow 
\sigma_2s_5.$$

A union of \re{dfsgdsfhdghdghdf555}
and \re{sdfgdfsgds345264435}
is a neighborhood 
with five leaves, 
such that \bi\i
two of there leaves are terminal, 
and their output term is equal to 1, 
and \i other leaves 
are non-terminal, and
each of them is included in 
some its ancestor.\ei

Thus, the union of neighborhoods
\re{dfsgdsfhdghdghdf555}
and \re{sdfgdfsgds345264435},
with the above
embeddings of non-terminal leaves,
is a SD of FP $\Sigma'(\Sigma)$.

On the reason of theorem \ref{th3}
we conclude that equality
 \re{dsfasdgdfgdsfgds} holds.
 $\blackbox$

\section{Conclusion}

In the article, we have introduced the concept of a state diagram of a functional program and have proposed a verification method based on the concept of a state diagram.

The main advantage of the proposed verification method is the possibility of its full automation: a construction of a state diagram for a functional program can be performed automatically using a fairly simple algorithm.

One of the problems for further research related to the concept of a state diagram is the following: to find a sufficient condition (possibly the strongest) for a functional program, such that if a functional program satisfies this condition, then it has a state diagram.

\end{document}